\documentclass[reprint, amsmath, amssymb, aps, nofootinbib, onecolumn, superscriptaddress]{revtex4-1}

\usepackage{graphicx}  % needed for figures
\usepackage{dcolumn}   % needed for some tables
\usepackage{bm}        % for math
\usepackage{amssymb, amsmath}
\usepackage{subcaption}
\usepackage{booktabs,tabularx}
\usepackage[dvipsnames]{xcolor}
\usepackage{float}
\usepackage{upgreek}
\usepackage{multirow}
\usepackage[nolist,nohyperlinks]{acronym}
\usepackage{relsize}
\usepackage{xfrac}
\usepackage{mathtools}
\usepackage{colortbl}
\usepackage[export]{adjustbox}
\usepackage{marginnote}
\usepackage{colortbl}
\usepackage{hhline}
\usepackage{ulem}

\usepackage[latin1]{inputenc}
\usepackage[francais]{babel}
\definecolor{blue}{rgb}{0.36, 0.54, 0.66}
\definecolor{amaranth}{rgb}{0.9, 0.17, 0.31}
\definecolor{pink}{rgb}{0.87, 0.56, 0.81}
\definecolor{ao}{rgb}{0.0, 0.5, 0.0}
\definecolor{maroon}{rgb}{0.76, 0.13, 0.28}
\definecolor{cardinal}{rgb}{0.77, 0.12, 0.23}
\definecolor{lightcardinal}{rgb}{0.97, 0.42, 0.53}
\definecolor{frenchlila}{rgb}{0.53, 0.38, 0.56}
\definecolor{yellow}{rgb}{1.0, 1.0, 0.87}
\definecolor{lightseagreen}{rgb}{0.7, 0.92, 0.68}
\definecolor{gray}{rgb}{0.9, 0.9, 0.9}
\definecolor{lightblue}{rgb}{0.66, 0.84, 0.96}

\usepackage[pagebackref=false,
			hyperindex=true,
			citecolor=cardinal,
			colorlinks=true,
			linkcolor=blue,
			urlcolor=blue]{hyperref}

\newcommand\mnras{MNRAS}
\newcommand{\jcap}{J. Cosmol. Astropart. Phys.}

\makeatletter
    \def\CT@@do@color{%
      \global\let\CT@do@color\relax
            \@tempdima\wd\z@
            \advance\@tempdima\@tempdimb
            \advance\@tempdima\@tempdimc
    \advance\@tempdimb\tabcolsep
    \advance\@tempdimc\tabcolsep
    \advance\@tempdima2\tabcolsep
            \kern-\@tempdimb
            \leaders\vrule
                    \hskip\@tempdima\@plus  1fill
            \kern-\@tempdimc
            \hskip-\wd\z@ \@plus -1fill }
\makeatother
    
\begin{document}

\title{The Laser Interferometer Space Antenna mission in Greece \\ White Paper\footnote{The shape and organization of this document, have been inspired by the \href{https://meetings.triumf.ca/event/220/attachments/2327/2641/LISA_Canada_White_Paper_2021_circulation_v4.pdf}{work of our Canadian Colleagues}.}
}

\author{Nikolaos  Karnesis}
\affiliation{Department of Physics, Aristotle University of Thessaloniki, Thessaloniki 54124, Greece}
\author{Nikolaos Stergioulas}
\affiliation{Department of Physics, Aristotle University of Thessaloniki, Thessaloniki 54124, Greece}
\author{George Pappas}
\affiliation{Department of Physics, Aristotle University of Thessaloniki, Thessaloniki 54124, Greece}
\collaboration{Workshop and white paper coordinators}

%%%%%

\author{Charis Anastopoulos}
\affiliation{Department of Physics, University of Patras, 26504, Patras, Greece}

\author{John Antoniadis}
\affiliation{Institute of Astrophysics, Foundation for Research and Innovation - Hellas, N. Plastira 100, 70013, Heraklion, Greece}

\author{Theocharis Apostolatos}
\affiliation{Department of Physics, National and Kapodistrian University of Athens, Panepistimiopolis Zografos, 15783, Athens, 
Greece}

\author{Spyros Basilakos}
\affiliation{National Observatory of Athens, Lofos Nymfon, 11852 Athens, Greece}
\affiliation{Academy of Athens, Research Center for Astronomy and Applied 
Mathematics,\\
Soranou Efesiou 4, 11527, Athens, Greece}
\affiliation{
School of Sciences, European University Cyprus, Diogenes Street, Engomi 1516 
Nicosia}

\author{Kyriakos Destounis}
\affiliation{Theoretical Astrophysics, IAAT, University of T\"ubingen, Germany}

\author{Eleni Areti}
\affiliation{Department of Physics, National and Kapodistrian University of Athens, Panepistimiopolis Zografos, 15783, Athens, 
Greece}

\author{Georgios Lukes-Gerakopoulos}
\affiliation{Astronomical Institute of the Czech Academy of Sciences, Bo\v{c}n\'{i} II 1401/1a, CZ-141 00 Prague, Czech Republic}  

\author{Konstantinos N. Gourgouliatos}
\affiliation{Department of Physics, University of Patras, 26504, Patras, Greece}  

\author{Kostas D. Kokkotas}
\affiliation{Theoretical Astrophysics, IAAT, University of T\"ubingen, Germany}  

\author{George Kottaras}
\affiliation{SPACE ASICS S.A., Athens, 10640, Greece}

\author{V.K. Oikonomou}
\affiliation{Department of Physics, Aristotle University of Thessaloniki, Thessaloniki 54124, Greece} 

\author{Theodoros Papanikolaou}
\affiliation{National Observatory of Athens, Lofos Nymfon, 11852 Athens, 
Greece}

\author{Leandros Perivolaropoulos}
\affiliation{Department of Physics, University of Ioannina, 45110, Ioannina, Greece}

 \author{Manolis Plionis}
\affiliation{National Observatory of Athens, Lofos Nymfon, 11852 Athens, 
Greece}
\affiliation{Department of Physics, Aristotle University of Thessaloniki, Thessaloniki 54124, Greece}
\affiliation{CERIDES, Center of Excellence in Risk \& Decision Sciences, European University of Cyprus, Cyprus}

 \author{Emmanuel N. Saridakis}
\affiliation{National Observatory of Athens, Lofos Nymfon, 11852 Athens, 
Greece}
\affiliation{Department of Astronomy, School of Physical Sciences, University of 
Science and Technology of China, Hefei, Anhui 230026, China}

\author{Theodoros Sarris}
\affiliation{Department of Electrical and Computer Engineering, Democritus University of Thrace, Xanthi 67132, Greece}

\author{Elias~C.~Vagenas}
\affiliation{Department of Physics, Kuwait University, P.O. Box 5969, Safat 13060, Kuwait.}

\author{Wolf von Klitzing}
\affiliation{Institute of Electronic Structure and Laser, Foundation for Research and Technology-Hellas (FORTH-IESL),Heraklion 70013, Greece}

\date{\today}% It is always \today, today,

%
%
%
%		ABSTRACT
%
%
%
\begin{abstract}
The Laser Interferometer Space Antenna (LISA) mission,  scheduled for launch in the mid-2030s, is a gravitational wave space observatory designed to detect sources emitting in the millihertz band. LISA is an ESA flagship mission, currently entering the Phase B development phase. It is expected to help us improve our understanding of our Universe by measuring gravitational wave sources of different types, with some of the sources being at very high redshifts ($z\sim 20$). On the 23$^{\rm rd}$ of February 2022 we organized the 1$^\mathrm{st}$ {\it LISA in Greece Workshop}. This workshop aimed to inform the Greek scientific and tech industry community about the possibilities of participating in LISA science and the LISA mission, with the support of the Hellenic Space Center (HSC). In this white paper, we summarize the outcome of the workshop, the most important aspect of it being the inclusion of $15$ Greek researchers to the LISA Consortium, raising our total number to $22$. At the same time, we present a road-map with the future steps and actions of the Greek Gravitational Wave community with respect to the LISA mission. 
\vspace{.5cm}\newline
\begin{center}
    {\large \url{https://indico.physics.auth.gr/e/lisa_gr_2022}}
\end{center}
% \vspace{.1cm}
% \begin{center}
%     \includegraphics[width=0.33
% \textwidth,valign=c]{LISAConstellation.pdf}
%\end{center}

\end{abstract}

\maketitle

%
%
%
%		EXECUTIVE SUMMARY
%
%
%
\section*{Executive summary
\label{sec:execsummary}}

The LISA space mission will open a new window to the Universe by detecting gravitational waves in the $\mathrm{mHz}$ bandwidth. LISA is going to be the first space-borne Gravitational Wave observatory, to be launched in the mid-2030s. In February 2022, Greek researchers organized the \href{https://indico.physics.auth.gr/event/11/}{1$^\mathrm{st}$ {\it LISA in Greece Workshop}} with the aim of informing and coordinating the local researchers and industrial companies in contributing to the LISA mission. During the Workshop, several speakers from the European Space Agency (ESA) and the LISA Consortium gave an overview of ongoing LISA science activities and the mission-development status.

\paragraph{The Greek Research Community:} At the time of the workshop, only one research group (at the University of Thessaloniki) in Greece were members of the LISA Consortium (participating in the data analysis effort). One of the workshop's main objectives was to attract new members to the LISA Consortium and determine the possibilities of further participation in the LISA mission. In this regard, the workshop was quite successful:
\begin{itemize}
    \item A total of 139 persons from the local academia and industry registered for the meeting. 
    \item  The workshop clearly demonstrated that there already exists a high level of expertise on LISA-related topics among the Greek scientific and space-industry communities. Participants reviewed their recent work on a broad range of disciplines such as gravitational-wave astronomy, cosmology, extreme stellar environments and stellar astrophysics, galaxy formation, particle physics and dark matter.
    \item In addition, it was shown that the LISA Consortium (and the mission in general) is open to new members, and has the potential to support the expertise across Greece. On the other hand, Greek scientists are prepared to step up and assume their corresponding roles within the Consortium. At the same time, more than 70 high-tech companies are active in Greece in Space-related products, technologies and services, engaging over 2500 employees, of which the vast majority constitutes highly educated personnel in Space technology and applications. With a total turnover of nearly 200 million euros and a growth rate of 11\% per year (well over the corresponding global rate of 6.7\%), the Greek Space Industry is a health and expanding field, with strong participation in numerous European Space Programs. Acknowledging this expanding activity and growing interest over the last years, in 2019 the Greek Government tripled the country's investment in the mandatory and optional ESA programs. More specifically, Greece invested 33 million euros in the optional and 51 million euros in the mandatory ESA programs for the programmatic period 2020-2022 and 2020-2024, respectively.
\end{itemize}

\paragraph{Coordinating with Space Agencies:} During the workshop, representatives from the ESA and LISA Consortium gave an overview of the LISA mission, focusing on both the instrument, and the scientific potential of the LISA data. The presenters also described the different Working Groups, and their function within the Consortium.  Most importantly, a line of communication between the workshop organizers, ESA, and the Hellenic Space Center (HSC) was established. This novel network will support the determination of the different opportunities for Greece to the development of the mission. These opportunities include the development of flight hardware (e.g. optics), computing infrastructure (hardware and software), and science. The workshop also proved that Greece has considerable experience and expertise in all the aforementioned areas.
\paragraph{Planning for our next steps:} Analyzing the workshop outcome, Greek researchers expressed interest in participating in the LISA Consortium, as well as in collaborating on proposals aimed to fund LISA-related research. The interested parties recognized the scientific potential of the newly founded Greek network, which will open new possibilities for collaboration with experts around the globe, enable student exchanges, allow for training in interdisciplinary research fields, which will strengthen the local science and technology community. 

\newpage

%
%
%
%		LISA mission
%
%
%
\section{The LISA mission
\label{sec:lisa}}
\begin{figure*}[h!]
\includegraphics[width=0.45
\textwidth,valign=c]{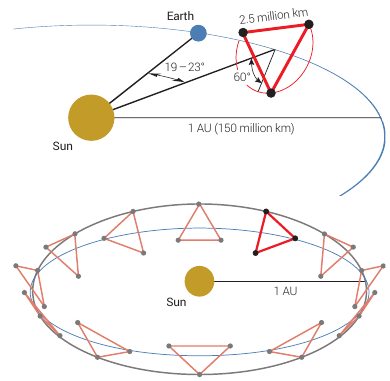}% Here is how to import EPS art
\includegraphics[width=0.55
\textwidth,valign=c]{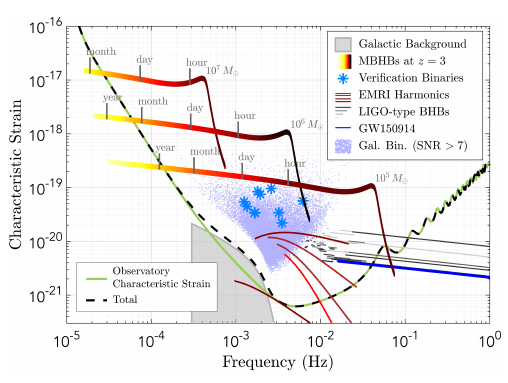}
\caption{\label{fig:lisa} {\it Left}: Cartoon of the ESA LISA constellation and the orbits designed for the mission. {\it Right}: Plot of the LISA sensitivity and the most probable sources that LISA can probe. We expect to measure supermassive and stellar-mass black hole binaries, as well as the ensemble signal originating from the ultra-compact binaries of our Galaxy (mostly Double White Dwarfs). Taken from~\cite{2017arXiv170200786A}.}
\end{figure*}

LISA is a space-based Gravitational Wave (GW) observatory mission scheduled for launch in the mid-2030s~\cite{2017arXiv170200786A, scird} (figure~\ref{fig:lisa}). LISA is going to probe the $\mathrm{mHz}$ range of the GW spectrum, aiming to measure signals from a variety of sources.  These include supermassive black hole binaries~\cite{PhysRevLett.96.111102, berti2006}, stellar-mass black hole Binaries~\cite{Babak2023lro,ligocatalogue, PhysRevLett.118.121101, PhysRevLett.122.081102, PhysRevD.91.022003, sesanasobbhs, Abbott2020mjq}, ultra-compact binaries originating in our Galaxy~\cite{Hiscock_2000, Kalogera_2001, Nelemans2001hp, Marsh_2011}, Extreme Mass Ratio Inspirals~\cite{PhysRevD.94.124042, PhysRevD.86.104050, Chua_2015, Barack2005aj}, and a Stochastic GW Background that may originate from cosmological sources~\cite{Caprini_2018, PhysRevD.79.062002} (see right panel of figure~\ref{fig:lisa}). Measuring such signals will have an immense impact on GW physics, Cosmology and Astrophysics~\cite{2017arXiv170200786A}.  

LISA will be comprised by a constellation of three spacecrafts on heliocentric orbit. Each spacecraft will enclose two test masses kept in free-fall conditions. The differential distance between the test masses of far-away spacecrafts (across $2.5\times10^6$~$\mathrm{km}$) will be monitored by means of laser interferometry. Since the arms of this type of interferometric detectors are not fixed, the data streams have to be combined on ground by employing time delayed interferometry in order to cancel the laser noise~\cite{tdi, aet}. The concept of space-borne GW observatories was tested with the ESA LISA Pathfinder (LPF) mission, which was a technology demonstrator for LISA~\cite{lpf_prl}. The success of LPF, together with the recent advances in GW astronomy, paved the way for such future space-based observatories.

Unlike GW signals observed with current ground detectors, LISA sources are expected to be long-lived, with thousands of strong GW signatures present in the data overlapping in time and/or in frequency. This implies that all GW signals from various sources will have to be fitted for and characterized simultaneously. The number and density of GW sources in the LISA band presents a data analysis challenge. A scheme to tackle this challenge has already been proposed, and is referred to as \emph{Global Fit}~\cite{Robson_2017}, which focuses on simultaneously detecting and classifying overlapping signals. Thus, within the LISA Consortium, there has been great effort invested in the data analysis techniques to be used during and after mission operations.  

Currently, LISA development is on the beginning of the {\it phase B} stage, where the initial design is being defined and communicated with ESA. We believe that this offers a great opportunity for the scientific community and tech industry based in Greece, to actively contribute to the development of the mission as parts of a consolidated national effort supported by the Hellenic Space Center (HSC). A contribution may come both in the form of instrumentation (e.g.\,optical flight hardware), as well as the data analysis and software products. Several spacecraft components are still under definition study, while the final construction assignment is expected to be solidified within the next couple of years. On the other hand, such an effort would benefit from increased coordination and synergies as well as additional investments in person power, training and infrastructure (for instance in the form of dedicated computing resources to tackle data-analysis challenges by the time that LISA commences operations. 

%
%
%
%		Opportunities for Greek involvement in LISA 
%
%
%
\section{Opportunities for further involvement of Greece in LISA
\label{sec:greekoportinities}}

The LISA mission is organized following two main branches. The main branch is the European Space Agency (ESA), and the second branch is the LISA Consortium. ESA is responsible for the development and management of the mission, which means that is also responsible for managing and delivering the three spacecrafts, the launch vehicle, the ground support, and many components of the instrument hardware. The Consortium on the other hand is comprised mostly by academics around the globe, and is providing the scientific support to ESA. The support provided by the Consortium comes as science deliverables (Astrophysics, Cosmology, Fundamental Physics) but also as data analysis and data processing products (software, algorithms, computing power, calibration, etc) as well as instrumentation (various components of the main science measurement hardware). During the workshop, all these points were presented and discussed between the participants and invited speakers. 

\paragraph{\bf Potential hardware opportunities}

At the heart of LISA lies an ultra-stable optical interferometer, for which a unique optical bonding technology has been developed.  The Space-Optics-Laboratory at FORTH-IESL has developed unique knowledge in highly-stable optics for laser conditioning and distribution. This was developed through two ESA contracts, for atom-interferometry in space, where the requirements of extreme extinction ratios and frequency stability required fiber-coupled breadboards. This  technology might find its use in LISA, e.g. in laser conditioning. 
Another area, where Greece has acquired some expertise, is the analysis, generation and control of radio-frequency signals. 

Key components of the scientific payload of LISA, such as the LISA phase-meter, are Analogue to Digital Converters (ADCs), which need to be of high speed and high performance, while having stringent power dissipation (thermal) requirements. SPACE-ASICS (SA) and Democritus University of Thrace (DUTH) personnel have acquired their pioneering expertise in the development of radiation-hardened (RH), low power, space qualified mixed Analog/Digital ASICs, through their over 20-year long participation in Advanced Technology ESA projects and a series of NASA Missions\footnote{These include the ETM-ASIC, a RH Essential Telemetry ASIC with a 12-bit ADC, the MCI-ASIC, a mixed signal Analog Front End ASIC for Multichannel (16 channels) Data Acquisition at 20 Msps, and a prototype high-resolution (16-bit) multichannel (8 channel) ASIC at 20 Msps.}. Related to ASIC developments, SA and DUTH have extensive experience in the testing, verification, qualification and characterization campaigns of ASICs, including thermal and radiation testing, which are essential steps in the development of a space mission's components\footnote{Radiation hardened is achieved by design, reaching a TID of up to 1 Mrad and SEE (Single Event Effect) levels such as: SEL immune up to 82 MeV/mg/cm2 and SEU free up to 57 MeV/mg/cm2, and qualification and testing have been done under ESA}. 

In conclusion, even though the design of LISA is in an advanced state, we believe that Greek companies and research institutes might still have a role to play in the development and construction of hardware for LISA, as well as in testing and qualification campaigns.

\paragraph{\bf Data processing and simulations.}

As already discussed, LISA is going to be a signal-dominated observatory. This means, that we will have to overcome significant challenges concerning the analysis of the data. Thus, a lot of effort is being invested in the solution of the future data analysis challenges for LISA. The LISA Data Processing Group (LDPG), together with the LISA Science Group (LSG) are the Consortium Working Groups that are coordinating this effort\footnote{The structure of the Consortium has recently started changing, due to entering the Phase B of mission development. Now different {\it Experts Groups} are being shaped, in order to contribute to more advanced studies of the mission.}. In particular, the LDPG coordinated major activities, such as simulations (science and instrument performance simulators), and also the prototyping of the initial noise reduction pipeline. In particular, a lot of effort is being invested in developing an accurate simulator of the LISA mission, an activity that is being coordinated by the Simulation Group, which is part of the LDPG. The Simulation Group coordinates the development of the several modular components that together are combined to build an accurate simulator of the mission. In particular, the modules under development are focusing on the orbits of the constellation, the instrumental noises and the instrument calibration, the time delay interferometry algorithms~\cite{tdi, aet}, and finally the response of the antenna to the GW signals. A great portion of this work, especially for the case of the models of the instrumental noises, has been inherited by the LISA Pathfinder mission~\cite{lpf_prl,lpf_prl2,lpf_cal,glitches,optimaldes}. Finally, it provides support for different Consortium needs (organizational and/or tools for computing scientific figures of merit of the mission), and most importantly, the design of the Distributed Data Processing Centres.

The \href{https://lisa-ldc.lal.in2p3.fr/}{LISA Data Challenge Group} (LDC), together with the LDPG, plays a major role in the development of the mission. At the same time, it contributes to the training of our scientific personnel and testing our data analysis algorithms and pipelines to realistic data. Thus, it will significantly contribute to improve our abilities to extract the full scientific potential from the LISA mission data. The LDC provides a common data analysis framework, by maintaining waveform conventions, and performing realistic simulations that aim to solve different types of challenges. All of this work is performed in close collaboration with the LDPG, the Astrophysics, the Cosmology, and the Waveform Working Groups. Until the writing of this document, there have been two data challenges released. The first challenge, the \href{https://lisa-ldc.lal.in2p3.fr/challenge1}{``Radler'' or LDC1} focused on building the community by introducing basic problems in LISA data analysis. The second challenge, \href{https://lisa-ldc.lal.in2p3.fr/challenge2a}{LDC2a} and \href{https://lisa-ldc.lal.in2p3.fr/challenge2b}{LDC2b} introduces the more realistic scenarios of LISA data analysis, where multiple types of GW sources are overlapping, and the statistical properties of the instrumental noise are departing from our ideal Gaussian assumptions.

The LDPG and LDC groups have undertaken the challenging and essential role of preparing for analyzing the very complicated future LISA data. With this in mind, there is critical need for manpower and new ideas for data analysis algorithms to aid to this purpose. Thus, there are considerable opportunities for Greek researchers to join and contribute to this effort. It is worth noting here, that already the \href{https://niksterg.github.io/gw-group/}{AUTh Group} is contributing in this direction, by developing a proposal for a data analysis pipeline funded by the ESA PRODEX program~\cite{Karnesis2023ras}, and also participating in different data-driven projects withing the Consortium. For example, there is a strong need to develop data analysis techniques that are robust against data artefacts, such as gaps~\cite{auth5_gaps,auth9_gap2} and glitches~\cite{glitches,Sasli2023mxr}. Another example is the important study requested by the Consortium to probe the mission science performance depending on the mission duration~\cite{auth4_missdur}. At the same time, the LISA data analysis requires an abundance of computational resources, which also means that the development of a LISA Computational Center in Greece is a viable and attractive option. Additionally, in recent years there has been an exponentially growing scientific computing community in Greece, as well as valuable expertise that originates from both the academic and the industrial sector. Thus, there are ample possibilities for Greek researchers to actively contribute in all aspects of the LISA data analysis and data processing.

\paragraph{\bf Astrophysics, Cosmology, and Fundamental Physics}

LISA, being sensitive in the millihertz band, will expand the current GW observational window, opening more avenues of scientific exploration that cover a much broader scope for doing Astrophysics as well as Fundamental Physics. LISA sources will comprise a broad and diverse group of new and old GW emitters, including familiar objects such as stellar origin BHs in early stage merging binaries, or so far unobserved with GWs objects such as massive BHs in similar mass, extreme mass-ratio and intermediate mass-ratio binaries, galactic binaries, primordial black holes or even more exotic objects such as cosmic strings, sources of cosmological origin, or beyond standard model and beyond GR objects, just to name some examples. Furthermore, some of these sources will be observable across GW bands providing the opportunity for multi-band observations while other sources will be visible across observational windows (GW, E/M, or other) galvanizing the field of multi-messenger Astrophysics. The diversity of potential sources and the pluralism of physics involved, provides the opportunity for researchers from many sub-disciplines to be involved, including Astrophysics, Gravitational Physics, Cosmology, and Particle Physics, thus opening the door to the broader Greek scientific community, which includes the strong communities of High Energy Physics and broader Astrophysics, for being involved with LISA science. The {\it LISA in Greece Workshop} witnessed the expression of interest from the wider Greek scientific community, where researchers from almost all Greek Universities and relevant research Institutes presented their interest in the corresponding scientific goals of the mission. 

Each presenter focused on different aspects of the LISA science, focusing on different types of GW sources that can be measured with LISA. A prime example is the science associated with Extreme Mass Ratio Inspirals (EMRIs). EMRIs are binary sources composed of an orders-of-magnitude lighter compact object, such as a black hole or a neutron star, inspiraling into a supermassive black hole. During the inspiral the secondary object follows complicated adiabatically-damped orbits, which allow us to map the spacetime geometry of the supermassive black hole to unprecedented accuracy and test our theoretical predictions regarding Fundamental Physics \cite{Sotiriou05,Glampedakis13,Eleni20,Destounis.2020kss,Timogiannis21}. Especially, the passage through a resonance between the corresponding fundamental orbital frequencies of the inspiraling low-mass object, is expected to reveal the exact nature of the central massive object~\cite{Apostolatos09,Destounis.2021mqv,Destounis.2021rko}. 

Another example is the GW sources that can potentially be measured with optical means as well. These could be binary magnetized sources, such as double neutron stars, that have magnetospheres and winds which interact weakly with each other, leading to modulations of the pulsar signal \cite{2005MNRAS.362.1078L, 2012ApJ...750..130P}. Prior to merger, such interactions should become stronger and provide electromagnetic counterparts to early gravitational waves, either in the form of shattering flares or strongly interacting and reconnecting magnetospheres \cite{2013ApJ...777..103T,2019MNRAS.482.1942G}. Even isolated neutron stars, may be sources of observable gravitational waves. Deformations in the form of mountains \cite{2021MNRAS.502.4680S} arising from magnetic fields misaligned with the rotation axis, or tangled ones \cite{2018ApJ...852...21G,2020MNRAS.495.1692G,2021ApJ...909..101I} provide the essential multipolar moments generating gravitational waves, whose amplitudes are expected to be within the observing capabilities of LISA. In addition, one of the most abundant sources of GWs in the LISA band will be the ultra-compact binaries in the vicinity of our own Galaxy~\cite{2017arXiv170200786A,gal1,gal2,gal3}. These are mostly comprised by Double White Dwarf binaries, and will act as multi-messenger laboratories that will help us probe the properties of our Galaxy~\cite{auth1_geor,auth2_triplets,auth3_obs}, and also perform tests of General Relativity~\cite{gal4}. LISA will be able to resolve a small percentage $\sim\mathcal{O}(10^4)$ of their total population ($\mathcal{O}(10^6)$), while the rest will generate an astrophysical stochastic confusion signal that will dominate the LISA band between $10^{-4}$ and $2\times10^{-3}~\mathrm{Hz}$~\cite{auth6_gal}. 

The most profound task of LISA will be the observation of a stochastic primordial gravitational wave background, which if observed will be mainly a smoking gun for inflationary theories, although there exist additionally more exotic scenarios that can generate stochastic primordial gravitational waves. LISA will probe truly primordial modes which became subhorizon near the end and during the early stages of the radiation domination era, during the reheating era. These modes cannot be probed by CMB experiments because the CMB modes acquire non-linear character for wavelengths below 10 Mpc. Thus, LISA will offer the possibility of probing the early Universe in a wide range of frequencies. Therefore, it will offer the possibility of testing several theories of pure general relativity and of modified gravity via the measured energy spectrum \cite{oikonomou1,oikonomou2,oikonomou3}. Plausible and exciting scenarios, such as the presence and absence of a signal at specific frequencies, can also give important information for the physics of the Universe at the second horizon crossing post-inflationary. 

On the other hand, one of the most important task of modern Cosmology, is to measure the value of the Hubble constant, which will reveal the history and evolution of our Universe. Especially nowadays, great effort has been invested into solving the 'Hubble tension'~\cite{2021CQGra..38o3001D,2021arXiv210505208P,2022arXiv220306142A}. Most of the theoretical approaches for the resolution of this tension have difficulty to reconcile with other cosmological probes~\cite{2021CQGra..38o3001D,2021MNRAS.504.3956A}. A novel approach which appears to be consistent with both geometric and dynamical cosmological probes involves a rapid transition of the absolute luminosity of type Ia supernovae (SnIa) to a value lower by about $10\%$ occurring within the last $50$ to $150~\mathrm{Myrs}$ (redshift $z<0.01$) \cite{2021PhRvD.104b1303M,2021PhRvD.103h3517A,2022PhRvD.105f3538A}. Such a transition could be induced by a sudden decrease of the gravitational constant for objects observed beyond a critical distance \cite{2022arXiv220310374P,2021PhRvD.104l3511P} about 20-40 Mpc (light emitted from them 50-150Myrs ago) which is consistent (or even favored in some cases) with current astrophysical probes \cite{2022PhRvD.105f3538A,2021Univ....7..366A,2021PhRvD.104l3511P,2022arXiv220105846A} and with solar system chronology \cite{2022arXiv220108997P}. 

Gravitational waves offer an independent probe of fundamental physics \cite{2020GReGr..52...81B,2017JCAP...05..031C,2019CQGra..36n3001B}  and in particular for the measurement of both the Hubble constant \cite{2022SCPMA..6559811Z,2017PhRvD..95h3525K,2021arXiv211200241T} and the gravitational constant \cite{2019JCAP...07..024B,2018PhRvD..98b3510B} values, applicable for the coalescing compact binary systems that emitted them. Within 5 years of operation time, LISA is expected to measure the Hubble constant with an $1\%$ accuracy \cite{2022SCPMA..6559811Z} using dark and bright standard sirens which are self calibrated from gravitational considerations and independent of SnIa standard candle and standard ruler calibration~\cite{2021PhRvD.103h3526S}. This accuracy level is better than most present methods and has a comparable accuracy to that expected from other future space missions. In addition, constraints on a possible change of the gravitational constant (Planck mass) will be imposed at a level of $1\%$ using the anticipated modification of the Hubble expansion friction term in the evolution of gravitational waves ~\cite{2019JCAP...07..024B,2018PhRvD..98b3510B}. Thus, the LISA mission will play a critical role in ruling out or confirming this class of models for the resolution of the Hubble tension while also constraining a wide range of modified gravity models that predict evolution of the strength of the gravitational interactions.

The LISA mission can also immensely contribute in the improvement of our understanding behind the Primordial Black Hole (PBH) physics~\cite{1967SvA....10..602Z, Carr1974nx,1975ApJ...201....1C, Chapline1975ojl,Clesse2017bsw} and the associated multidisciplinary physics research fields through the associated to PBHs GW signals~\cite{Nakamura1997sm, Ioka1998nz, Eroshenko2016hmn, Raidal2017mfl, Zagorac2019ekv,Hooper2020evu,Anantua2008am,Dong2015yjs,Domenech2021ztg,Papanikolaou2020qtd,Papanikolaou2022chm}. Very interestingly,  the PBH formation process is unavoidably connected to the emission of a stochastic gravitational-wave background (SGWB) at second order in cosmological perturbation theory, which for PBH masses around $\mathcal{O} (10^{-15}\div 10^{-8})~\mathrm{M}_\odot$ can be potentially detected by LISA~\cite{Bartolo2018evs, Cai, Papanikolaou2020qtd, Karnesis2019mph,Flauger2020qyi,Baghi2023qnq}. Therefore, the detection or not of such a signal will give us access to the shape and the non-Gaussian~\cite{Cai2018dig} nature of the primordial curvature power spectrum which gave rise to PBHs on very small scales,  which are otherwise poorly constrained by other cosmological and astrophysical probes like CMB and LSS. This SGWB can give us access to early universe physics up to GUT scales~\cite{Papanikolaou2022did,Basilakos2023xof} and can serve as a novel probe constraining the underlying gravity theory~\cite{Papanikolaou2021uhe,Papanikolaou2022hkg}. Finally, with the LISA mission one can detect as well the SGWB from early and late PBH binaries as well as GWs from intermediate-mass binaries, from extreme mass ratio inspirals as well as from high redshift mergers shedding light in this way on the clustering properties of PBHs and explaining merging events within the mass gap $(2.5\div 5)\mathrm{M}_\odot$~\cite{LISACosmologyWorkingGroup2022jok} giving the possibility at the same time to detect for the first time sublunar primordial black holes of not astrophysical origin~\cite{LIGOScientific2020zkf}. 

The interaction of quantum systems with stochastic gravitational backgrounds may lead to observable decoherence phenomena, and in specific regimes, it could provide tests of theories of gravity as an emergent force \cite{AHGravDec, LagAn21}, or identify quantum gravity signals \cite{PWZ20}. The Patras group is part of the Science Definition Team of the Deep Space Quantum Link (DSQL) mission  of  JPL-NASA \cite{DSQL}, which aims to  establish quantum optical links across extremely long baselines through the Lunar Gateway moon-orbiting space station. The links will be employed in order to conduct fundamental physics experiments including long-range teleportation, tests of gravitational coupling to quantum systems, and advanced tests of quantum nonlocality. There are strong synergies and complementarity between the DSQL mission and LISA.

Additionally, gravitational waves, and especially those at the frequency range accessible by LISA, will offer crucial information on the allowed possible deviations from General Relativity. In particular,  modified theories of gravity typically predict various  effects on the gravitational waves properties (e.g. on their phase, amplitude, speed, polarization dispersion, damping, oscillations etc) \cite{CANTATA2021ktz,2022PrPNP.12503948A}. Therefore, LISA data will offer a novel tool to test General Relativity and detect possible smoking guns of modified gravity, and hence it will enlighten the discussion on the fundamental theory of gravity in Nature.

Finally, General Relativity (GR) and Quantum Mechanics (QM) are expected to intermingle at very small scales.  As already stated, the sources of the GWs to be detected by LISA are in the mHz range of GWs.  We would like to see if there will be deviations between the observational data and the theoretical predictions given by our standard theories, namely GR, LCDM, and Standard Model. If there are such deviations, then one may consider them as fingerprints of the yet-to-be constructed theory of Quantum Gravity (QG) and we would like to provide theoretical explanations in the framework of QG Phenomenology \cite{2008PhRvL.101v1301D, 2010PhRvL.104k9002D,2009CaJPh..87..233D,2011PhRvD..84d4013A}. Employing the generalized uncertainty principle (GUP) (see for instance \cite{2009PhLB..678..497A,2010PhLB..690..407D,2010PhLB..692..342D}) in the framework of QG Phenomenology, we can predict deviations or provide answers for already observed ones that take place in gravitational backgrounds such as those related to the physics of black holes \cite{2017PhLB..767..242S,2017EL....12040001V,2019PhRvD..99h4013V}, to the speed of gravitons and photons \cite{2021PhLB..81936429D}, or to Planck quantities \cite{2010JCAP...09..027B}, as well as deviations in cosmological backgrounds such as the Hubble tension \cite{2021EPJP..136..997A}, the  anomalous 21-cm signal at $z\simeq17.2$ \cite{2021arXiv211002340D}, the baryon asymmetry \cite{2022PhLB..82436841D,2021arXiv211101278D},  the post-inflation reheating \cite{2011JCAP...12..017C}. Furthermore, we would like to test the Generalized Uncertainty Principle (GUP) {\it per se} using  LISA. 
Of course, GR and QM do not reconcile and thus QG must occur. Therefore, LISA with its observations will be the best test bed for candidate theories of QG via GUP/QG effects.

%
%
%
%		CONCLUSION 
%
%
%
\section{Conclusion
\label{sec:summary}}

A national combined effort within the LISA community, would allow for a desirable know-how transfer, and fruitful collaborations with distinguished colleagues and research institutes. This would not only affect the competitiveness of Greek scientists and institutions, but the local tech industry as well. Apart from the direct consequences, such as grants and funding opportunities, our national participation to LISA would also build an important legacy for the years to come. LISA will inspire and involve young Greek scientists in gravitational-wave astronomy, an emerging and quickly developing multidisciplinary field. Last but not least, Greece would be officially a member of an ESA flagship mission, which aims to further advance our understanding of the Universe. 

We believe that the {\emph 1st LISA in Greece Workshop} highlighted all the aforementioned points, and it can be regarded as the starting point of further Greek involvement in the development of the mission. In particular, we list below the main outcomes of this meeting.
\begin{itemize}
    \item The Workshop gave an overview of LISA and its status, exposure to local experts, and informed the scientific and industry communities about the latest developments of the mission. 
    \item It established contact points between local researchers and industry to ESA and Consortium.
    \item We discovered possible collaborations between the local research experts.
    \item It increased participation to the LISA Consortium, by the creation of {\it four} new groups applying for LISA membership\footnote{At the time of the writing of this document, the four new LISA groups have submitted their membership applications to the LISA Consortium.}, with $15$ new scientists becoming associated with the LISA Consortium, raising our total number to 22\footnote{This figure includes both {\it Full} and {\it Associate} members.}.
    \item During the Workshop we explored the possibility of hardware contributions, both from academic labs, and from the industry, as well as the establishment of dedicated computing clusters.
    \item We explored the possibility of a well-organized Greek community, with the aim of further advancing the Greek participation to LISA. This may come in the form of personpower (students, postdocs, senior researchers) and software, computing or hardware development. 
\end{itemize}

\pagebreak

\begin{table*}[ht]
\small
\renewcommand{\arraystretch}{1.2}
\caption{\label{tab:workshop} The  schedule for the first {\it LISA in Greece} workshop. The morning session focused on LISA, both from the instrument and the science sides of the mission. The afternoon session focused on the local scientific groups and their work with gravitational-waves, as well as the capabilities of the local tech industry. An open discussion took place before the end of the meeting. For more details, visit \href{https://indico.physics.auth.gr/e/lisa_gr_2022}{\texttt{indico.physics.auth.gr/e/lisa\_gr\_2022}}.}
  \begin{center}
    \label{tab:table1}
    \begin{tabularx}{.98\textwidth}{|l | X | X |} % <-- Alignments: 1st column left, 2nd middle and 3rd right, with vertical lines in between
    \hline
      \rowcolor{gray}
       & \textbf{Talk Content/Speaker} & \textbf{Short description}\\
      \hline \hline 
      \cellcolor{lightcardinal}  & ``Welcome \& Introduction''
      \reversemarginpar\marginnote{\rotatebox{90}{\bf \color{cardinal} Introduction to the ESA LISA Mission}}[-.4cm] \newline\newline {\bf Speakers:} \newline Representative from the Ministry of \newline Digital Governance \newline Representative of AUTh authorities \newline I. Daglis (HSC) & Introduction \& scope of the workshop. \\ \hline\hline
      \cellcolor{lightcardinal} & ``Overview of the LISA mission: The science case of LISA'' \newline\newline {\bf Speaker:} \newline O. Jennrich (ESA) & Brief overview of the LISA mission and its scientific goals.\\ \hline\hline
      \cellcolor{lightcardinal}  & ``ESA \& the LISA mission'' \newline\newline {\bf Speaker:} \newline M. Gehler (ESA)  & The LISA mission from the ESA point of view. Timeline, organization, contributions, deliverables. \\
      \hline\hline 
      \cellcolor{yellow}  & ``LISA Data Analysis \& LISA Data Centers'' \reversemarginpar\marginnote{\rotatebox{90}{\bf \color{orange} Possible Contributions}}[.2cm] \newline\newline {\bf Speaker:} \newline A. Petiteau (APC/CEA) & Overview of the LISA Data Analysis challenges and requirements. Stating the problems, timeline, data products, and mission operations. Brief introduction to the LISA Consortium. Discussion on the Scope \& Requirements of the LISA data centers. Main structure and proposed outline. \\ \hline\hline
      \cellcolor{yellow} & ``The LISA Instrument'' \newline\newline {\bf Speaker:} \newline M. Hewitson (AEI) & Overview of the LISA instrument. List of possibilities for GR-based labs and industry. Brief introduction to the LISA Consortium. \\  \hline\hline
      \cellcolor{lightblue}  & ``Local expertise in Science (Astrophysics and Cosmology) and Space Missions (Instrumentation and Data Analysis)'' \reversemarginpar\marginnote{\rotatebox{90}{\bf \color{blue} Local Expertise \& Discussion}}[1cm] \newline\newline {\bf Speakers from:} \newline N. Stergioulas (AUTh), K. Vyrsokinos/D. Chatzitheoharis (AUTh), S. Basilakos, E. Saridakis (NOA), J. Antoniadis (UOC/FORTH), T. A. Apostolatos (UOA), C. Kouvaris (NTUA), L. Perivolaropoulos (UOI), K. Gourgouliatos (UOP), W. von Klitzing (IESL/FORTH), T. Sarris (DUTh) & Multiple speakers from our national institutes presenting their work and its relation to LISA and GWs in general, as well as laboratories with experience in space missions. For more details about this slot, please see the detailed program at \href{https://indico.physics.auth.gr/e/lisa_gr_2022}{\texttt{indico.physics.auth.gr/e/lisa\_gr\_2022}}. \\ \hline\hline
      \cellcolor{lightblue}  & ``Activities \& Perspectives of the Si-Cluster: the ESA-BIC Greece'' \newline\newline {\bf Speaker:} \newline Jorge Sanchez (Corallia/Si-Cluster) & ESA BIC Greece, a key element of the Greek Space Industry aimed to strengthen space related start-ups in Greece, has already started its operation. By supporting 25 incubatees over a period of 5 years, ESA BIC will underpin the creation, grow and strengthening of space activities in Greece, contribute to the national innovation and support transfer of technologies from/to the space sector.\\ \hline\hline
      \cellcolor{lightblue} & ``Open Discussion'' \newline\newline {\bf Chairs:} \newline N Stergioulas, N Karnesis, G Pappas (AUTh) \newline\newline
      {\bf Invited Participants:} \newline M Hewitson (AEI), A Petiteau (APC/CEA), M Gehler (ESA), O Jennrich (ESA), P McNamara (ESA), local institutes and industry.
      & Open discussions involving the scientific and tech industry participants. Further actions and future steps.\\ 
      \hline
    \end{tabularx}
  \end{center}
\end{table*}


\begin{thebibliography}{133}%
\makeatletter
\providecommand \@ifxundefined [1]{%
 \@ifx{#1\undefined}
}%
\providecommand \@ifnum [1]{%
 \ifnum #1\expandafter \@firstoftwo
 \else \expandafter \@secondoftwo
 \fi
}%
\providecommand \@ifx [1]{%
 \ifx #1\expandafter \@firstoftwo
 \else \expandafter \@secondoftwo
 \fi
}%
\providecommand \natexlab [1]{#1}%
\providecommand \enquote  [1]{``#1''}%
\providecommand \bibnamefont  [1]{#1}%
\providecommand \bibfnamefont [1]{#1}%
\providecommand \citenamefont [1]{#1}%
\providecommand \href@noop [0]{\@secondoftwo}%
\providecommand \href [0]{\begingroup \@sanitize@url \@href}%
\providecommand \@href[1]{\@@startlink{#1}\@@href}%
\providecommand \@@href[1]{\endgroup#1\@@endlink}%
\providecommand \@sanitize@url [0]{\catcode `\\12\catcode `\$12\catcode
  `\&12\catcode `\#12\catcode `\^12\catcode `\_12\catcode `\%12\relax}%
\providecommand \@@startlink[1]{}%
\providecommand \@@endlink[0]{}%
\providecommand \url  [0]{\begingroup\@sanitize@url \@url }%
\providecommand \@url [1]{\endgroup\@href {#1}{\urlprefix }}%
\providecommand \urlprefix  [0]{URL }%
\providecommand \Eprint [0]{\href }%
\providecommand \doibase [0]{http://dx.doi.org/}%
\providecommand \selectlanguage [0]{\@gobble}%
\providecommand \bibinfo  [0]{\@secondoftwo}%
\providecommand \bibfield  [0]{\@secondoftwo}%
\providecommand \translation [1]{[#1]}%
\providecommand \BibitemOpen [0]{}%
\providecommand \bibitemStop [0]{}%
\providecommand \bibitemNoStop [0]{.\EOS\space}%
\providecommand \EOS [0]{\spacefactor3000\relax}%
\providecommand \BibitemShut  [1]{\csname bibitem#1\endcsname}%
\let\auto@bib@innerbib\@empty
%</preamble>
\bibitem [{\citenamefont {Amaro-Seoane}\ \emph {et~al.}(2017)\citenamefont
  {Amaro-Seoane} \emph {et~al.}}]{2017arXiv170200786A}%
  \BibitemOpen
  \bibfield  {author} {\bibinfo {author} {\bibfnamefont {P.}~\bibnamefont
  {Amaro-Seoane}} \emph {et~al.},\ }\href@noop {} {\bibfield  {journal}
  {\bibinfo  {journal} {ArXiv e-prints}\ } (\bibinfo {year} {2017})},\ \Eprint
  {http://arxiv.org/abs/1702.00786} {arXiv:1702.00786 [astro-ph.IM]}
  \BibitemShut {NoStop}%
\bibitem [{\citenamefont {{LISA Science Study Team}}(2018)}]{scird}%
  \BibitemOpen
  \bibfield  {author} {\bibinfo {author} {\bibnamefont {{LISA Science Study
  Team}}},\ }\href
  {https://www.cosmos.esa.int/documents/678316/1700384/SciRD.pdf} {\emph
  {\bibinfo {title} {{LISA Science Requirements Document}}}},\ \bibinfo {type}
  {Tech. Rep.}\ \bibinfo {number} {{ESA-L3-EST-SCI-RS-001}}\ (\bibinfo
  {institution} {ESA},\ \bibinfo {year} {2018})\BibitemShut {NoStop}%
\bibitem [{\citenamefont {Baker}\ \emph {et~al.}(2006)\citenamefont {Baker},
  \citenamefont {Centrella}, \citenamefont {Choi}, \citenamefont {Koppitz},\
  and\ \citenamefont {van Meter}}]{PhysRevLett.96.111102}%
  \BibitemOpen
  \bibfield  {author} {\bibinfo {author} {\bibfnamefont {J.~G.}\ \bibnamefont
  {Baker}}, \bibinfo {author} {\bibfnamefont {J.}~\bibnamefont {Centrella}},
  \bibinfo {author} {\bibfnamefont {D.-I.}\ \bibnamefont {Choi}}, \bibinfo
  {author} {\bibfnamefont {M.}~\bibnamefont {Koppitz}}, \ and\ \bibinfo
  {author} {\bibfnamefont {J.}~\bibnamefont {van Meter}},\ }\href {\doibase
  10.1103/PhysRevLett.96.111102} {\bibfield  {journal} {\bibinfo  {journal}
  {Phys. Rev. Lett.}\ }\textbf {\bibinfo {volume} {96}},\ \bibinfo {pages}
  {111102} (\bibinfo {year} {2006})}\BibitemShut {NoStop}%
\bibitem [{\citenamefont {Berti}(2006)}]{berti2006}%
  \BibitemOpen
  \bibfield  {author} {\bibinfo {author} {\bibfnamefont {E.}~\bibnamefont
  {Berti}},\ }\href {\doibase 10.1088/0264-9381/23/19/S17} {\bibfield
  {journal} {\bibinfo  {journal} {Classical and Quantum Gravity}\ }\textbf
  {\bibinfo {volume} {23}} (\bibinfo {year} {2006}),\
  10.1088/0264-9381/23/19/S17}\BibitemShut {NoStop}%
\bibitem [{\citenamefont {Babak}\ \emph {et~al.}(2023)\citenamefont {Babak},
  \citenamefont {Caprini}, \citenamefont {Figueroa}, \citenamefont {Karnesis},
  \citenamefont {Marcoccia}, \citenamefont {Nardini}, \citenamefont {Pieroni},
  \citenamefont {Ricciardone}, \citenamefont {Sesana},\ and\ \citenamefont
  {Torrado}}]{Babak2023lro}%
  \BibitemOpen
  \bibfield  {author} {\bibinfo {author} {\bibfnamefont {S.}~\bibnamefont
  {Babak}}, \bibinfo {author} {\bibfnamefont {C.}~\bibnamefont {Caprini}},
  \bibinfo {author} {\bibfnamefont {D.~G.}\ \bibnamefont {Figueroa}}, \bibinfo
  {author} {\bibfnamefont {N.}~\bibnamefont {Karnesis}}, \bibinfo {author}
  {\bibfnamefont {P.}~\bibnamefont {Marcoccia}}, \bibinfo {author}
  {\bibfnamefont {G.}~\bibnamefont {Nardini}}, \bibinfo {author} {\bibfnamefont
  {M.}~\bibnamefont {Pieroni}}, \bibinfo {author} {\bibfnamefont
  {A.}~\bibnamefont {Ricciardone}}, \bibinfo {author} {\bibfnamefont
  {A.}~\bibnamefont {Sesana}}, \ and\ \bibinfo {author} {\bibfnamefont
  {J.}~\bibnamefont {Torrado}},\ }\href {\doibase
  10.1088/1475-7516/2023/08/034} {\bibfield  {journal} {\bibinfo  {journal}
  {JCAP}\ }\textbf {\bibinfo {volume} {08}},\ \bibinfo {pages} {034} (\bibinfo
  {year} {2023})},\ \Eprint {http://arxiv.org/abs/2304.06368} {arXiv:2304.06368
  [astro-ph.CO]} \BibitemShut {NoStop}%
\bibitem [{\citenamefont {Abbott}\ \emph {et~al.}(2018)\citenamefont {Abbott}
  \emph {et~al.}}]{ligocatalogue}%
  \BibitemOpen
  \bibfield  {author} {\bibinfo {author} {\bibfnamefont {B.~P.}\ \bibnamefont
  {Abbott}} \emph {et~al.} (\bibinfo {collaboration} {LIGO Scientific,
  Virgo}),\ }\href@noop {} {\  (\bibinfo {year} {2018})},\ \Eprint
  {http://arxiv.org/abs/1811.12907} {arXiv:1811.12907 [astro-ph.HE]}
  \BibitemShut {NoStop}%
%%CITATION = ARXIV:1811.12907;%%
\bibitem [{\citenamefont {Abbott}\ \emph {et~al.}(2017)\citenamefont {Abbott}
  \emph {et~al.}}]{PhysRevLett.118.121101}%
  \BibitemOpen
  \bibfield  {author} {\bibinfo {author} {\bibfnamefont {B.~P.}\ \bibnamefont
  {Abbott}} \emph {et~al.} (\bibinfo {collaboration} {LIGO Scientific
  Collaboration and Virgo Collaboration}),\ }\href {\doibase
  10.1103/PhysRevLett.118.121101} {\bibfield  {journal} {\bibinfo  {journal}
  {Phys. Rev. Lett.}\ }\textbf {\bibinfo {volume} {118}},\ \bibinfo {pages}
  {121101} (\bibinfo {year} {2017})}\BibitemShut {NoStop}%
\bibitem [{\citenamefont {Renzini}\ and\ \citenamefont
  {Contaldi}(2019)}]{PhysRevLett.122.081102}%
  \BibitemOpen
  \bibfield  {author} {\bibinfo {author} {\bibfnamefont {A.~I.}\ \bibnamefont
  {Renzini}}\ and\ \bibinfo {author} {\bibfnamefont {C.~R.}\ \bibnamefont
  {Contaldi}},\ }\href {\doibase 10.1103/PhysRevLett.122.081102} {\bibfield
  {journal} {\bibinfo  {journal} {Phys. Rev. Lett.}\ }\textbf {\bibinfo
  {volume} {122}},\ \bibinfo {pages} {081102} (\bibinfo {year}
  {2019})}\BibitemShut {NoStop}%
\bibitem [{\citenamefont {Aasi}\ \emph {et~al.}(2015)\citenamefont {Aasi} \emph
  {et~al.}}]{PhysRevD.91.022003}%
  \BibitemOpen
  \bibfield  {author} {\bibinfo {author} {\bibfnamefont {J.}~\bibnamefont
  {Aasi}} \emph {et~al.} (\bibinfo {collaboration} {LIGO Scientific
  Collaboration and Virgo Collaboration}),\ }\href {\doibase
  10.1103/PhysRevD.91.022003} {\bibfield  {journal} {\bibinfo  {journal} {Phys.
  Rev. D}\ }\textbf {\bibinfo {volume} {91}},\ \bibinfo {pages} {022003}
  (\bibinfo {year} {2015})}\BibitemShut {NoStop}%
\bibitem [{\citenamefont {Sesana}(2016)}]{sesanasobbhs}%
  \BibitemOpen
  \bibfield  {author} {\bibinfo {author} {\bibfnamefont {A.}~\bibnamefont
  {Sesana}},\ }\href {\doibase 10.1103/PhysRevLett.116.231102} {\bibfield
  {journal} {\bibinfo  {journal} {Phys. Rev. Lett.}\ }\textbf {\bibinfo
  {volume} {116}},\ \bibinfo {pages} {231102} (\bibinfo {year}
  {2016})}\BibitemShut {NoStop}%
\bibitem [{\citenamefont {Abbott}\ \emph
  {et~al.}(2020{\natexlab{a}})\citenamefont {Abbott} \emph
  {et~al.}}]{Abbott2020mjq}%
  \BibitemOpen
  \bibfield  {author} {\bibinfo {author} {\bibfnamefont {R.}~\bibnamefont
  {Abbott}} \emph {et~al.} (\bibinfo {collaboration} {LIGO Scientific,
  Virgo}),\ }\href {\doibase 10.3847/2041-8213/aba493} {\bibfield  {journal}
  {\bibinfo  {journal} {Astrophys. J. Lett.}\ }\textbf {\bibinfo {volume}
  {900}},\ \bibinfo {pages} {L13} (\bibinfo {year} {2020}{\natexlab{a}})},\
  \Eprint {http://arxiv.org/abs/2009.01190} {arXiv:2009.01190 [astro-ph.HE]}
  \BibitemShut {NoStop}%
\bibitem [{\citenamefont {Hiscock}\ \emph {et~al.}(2000)\citenamefont
  {Hiscock}, \citenamefont {Larson}, \citenamefont {Routzahn},\ and\
  \citenamefont {Kulick}}]{Hiscock_2000}%
  \BibitemOpen
  \bibfield  {author} {\bibinfo {author} {\bibfnamefont {W.~A.}\ \bibnamefont
  {Hiscock}}, \bibinfo {author} {\bibfnamefont {S.~L.}\ \bibnamefont {Larson}},
  \bibinfo {author} {\bibfnamefont {J.~R.}\ \bibnamefont {Routzahn}}, \ and\
  \bibinfo {author} {\bibfnamefont {B.}~\bibnamefont {Kulick}},\ }\href
  {\doibase 10.1086/312867} {\bibfield  {journal} {\bibinfo  {journal} {The
  Astrophysical Journal}\ }\textbf {\bibinfo {volume} {540}},\ \bibinfo {pages}
  {L5} (\bibinfo {year} {2000})}\BibitemShut {NoStop}%
\bibitem [{\citenamefont {Kalogera}\ \emph {et~al.}(2001)\citenamefont
  {Kalogera}, \citenamefont {Narayan}, \citenamefont {Spergel},\ and\
  \citenamefont {Taylor}}]{Kalogera_2001}%
  \BibitemOpen
  \bibfield  {author} {\bibinfo {author} {\bibfnamefont {V.}~\bibnamefont
  {Kalogera}}, \bibinfo {author} {\bibfnamefont {R.}~\bibnamefont {Narayan}},
  \bibinfo {author} {\bibfnamefont {D.~N.}\ \bibnamefont {Spergel}}, \ and\
  \bibinfo {author} {\bibfnamefont {J.~H.}\ \bibnamefont {Taylor}},\ }\href
  {\doibase 10.1086/321583} {\bibfield  {journal} {\bibinfo  {journal} {The
  Astrophysical Journal}\ }\textbf {\bibinfo {volume} {556}},\ \bibinfo {pages}
  {340} (\bibinfo {year} {2001})}\BibitemShut {NoStop}%
\bibitem [{\citenamefont {Nelemans}\ \emph {et~al.}(2001)\citenamefont
  {Nelemans}, \citenamefont {Yungelson},\ and\ \citenamefont
  {Portegies~Zwart}}]{Nelemans2001hp}%
  \BibitemOpen
  \bibfield  {author} {\bibinfo {author} {\bibfnamefont {G.}~\bibnamefont
  {Nelemans}}, \bibinfo {author} {\bibfnamefont {L.}~\bibnamefont {Yungelson}},
  \ and\ \bibinfo {author} {\bibfnamefont {S.~F.}\ \bibnamefont
  {Portegies~Zwart}},\ }\href {\doibase 10.1051/0004-6361:20010683} {\bibfield
  {journal} {\bibinfo  {journal} {Astron. Astrophys.}\ }\textbf {\bibinfo
  {volume} {375}},\ \bibinfo {pages} {890} (\bibinfo {year} {2001})},\ \Eprint
  {http://arxiv.org/abs/astro-ph/0105221} {arXiv:astro-ph/0105221} \BibitemShut
  {NoStop}%
\bibitem [{\citenamefont {Marsh}(2011)}]{Marsh_2011}%
  \BibitemOpen
  \bibfield  {author} {\bibinfo {author} {\bibfnamefont {T.~R.}\ \bibnamefont
  {Marsh}},\ }\href {\doibase 10.1088/0264-9381/28/9/094019} {\bibfield
  {journal} {\bibinfo  {journal} {Classical and Quantum Gravity}\ }\textbf
  {\bibinfo {volume} {28}},\ \bibinfo {pages} {094019} (\bibinfo {year}
  {2011})}\BibitemShut {NoStop}%
\bibitem [{\citenamefont {Berry}\ \emph {et~al.}(2016)\citenamefont {Berry},
  \citenamefont {Cole}, \citenamefont {Ca\~nizares},\ and\ \citenamefont
  {Gair}}]{PhysRevD.94.124042}%
  \BibitemOpen
  \bibfield  {author} {\bibinfo {author} {\bibfnamefont {C.~P.~L.}\
  \bibnamefont {Berry}}, \bibinfo {author} {\bibfnamefont {R.~H.}\ \bibnamefont
  {Cole}}, \bibinfo {author} {\bibfnamefont {P.}~\bibnamefont {Ca\~nizares}}, \
  and\ \bibinfo {author} {\bibfnamefont {J.~R.}\ \bibnamefont {Gair}},\ }\href
  {\doibase 10.1103/PhysRevD.94.124042} {\bibfield  {journal} {\bibinfo
  {journal} {Phys. Rev. D}\ }\textbf {\bibinfo {volume} {94}},\ \bibinfo
  {pages} {124042} (\bibinfo {year} {2016})}\BibitemShut {NoStop}%
\bibitem [{\citenamefont {Wang}\ \emph {et~al.}(2012)\citenamefont {Wang},
  \citenamefont {Shang},\ and\ \citenamefont {Babak}}]{PhysRevD.86.104050}%
  \BibitemOpen
  \bibfield  {author} {\bibinfo {author} {\bibfnamefont {Y.}~\bibnamefont
  {Wang}}, \bibinfo {author} {\bibfnamefont {Y.}~\bibnamefont {Shang}}, \ and\
  \bibinfo {author} {\bibfnamefont {S.}~\bibnamefont {Babak}},\ }\href
  {\doibase 10.1103/PhysRevD.86.104050} {\bibfield  {journal} {\bibinfo
  {journal} {Phys. Rev. D}\ }\textbf {\bibinfo {volume} {86}},\ \bibinfo
  {pages} {104050} (\bibinfo {year} {2012})}\BibitemShut {NoStop}%
\bibitem [{\citenamefont {Chua}\ and\ \citenamefont {Gair}(2015)}]{Chua_2015}%
  \BibitemOpen
  \bibfield  {author} {\bibinfo {author} {\bibfnamefont {A.~J.~K.}\
  \bibnamefont {Chua}}\ and\ \bibinfo {author} {\bibfnamefont {J.~R.}\
  \bibnamefont {Gair}},\ }\href {\doibase 10.1088/0264-9381/32/23/232002}
  {\bibfield  {journal} {\bibinfo  {journal} {Classical and Quantum Gravity}\
  }\textbf {\bibinfo {volume} {32}},\ \bibinfo {pages} {232002} (\bibinfo
  {year} {2015})}\BibitemShut {NoStop}%
\bibitem [{\citenamefont {Barack}\ and\ \citenamefont
  {Cutler}(2005)}]{Barack2005aj}%
  \BibitemOpen
  \bibfield  {author} {\bibinfo {author} {\bibfnamefont {L.}~\bibnamefont
  {Barack}}\ and\ \bibinfo {author} {\bibfnamefont {C.}~\bibnamefont {Cutler}}\
  }(\bibinfo {year} {2005})\ \Eprint {http://arxiv.org/abs/gr-qc/0502052}
  {arXiv:gr-qc/0502052} \BibitemShut {NoStop}%
\bibitem [{\citenamefont {Caprini}\ and\ \citenamefont
  {Figueroa}(2018)}]{Caprini_2018}%
  \BibitemOpen
  \bibfield  {author} {\bibinfo {author} {\bibfnamefont {C.}~\bibnamefont
  {Caprini}}\ and\ \bibinfo {author} {\bibfnamefont {D.~G.}\ \bibnamefont
  {Figueroa}},\ }\href {\doibase 10.1088/1361-6382/aac608} {\bibfield
  {journal} {\bibinfo  {journal} {Classical and Quantum Gravity}\ }\textbf
  {\bibinfo {volume} {35}},\ \bibinfo {pages} {163001} (\bibinfo {year}
  {2018})}\BibitemShut {NoStop}%
\bibitem [{\citenamefont {Regimbau}\ and\ \citenamefont
  {Hughes}(2009)}]{PhysRevD.79.062002}%
  \BibitemOpen
  \bibfield  {author} {\bibinfo {author} {\bibfnamefont {T.}~\bibnamefont
  {Regimbau}}\ and\ \bibinfo {author} {\bibfnamefont {S.~A.}\ \bibnamefont
  {Hughes}},\ }\href {\doibase 10.1103/PhysRevD.79.062002} {\bibfield
  {journal} {\bibinfo  {journal} {Phys. Rev. D}\ }\textbf {\bibinfo {volume}
  {79}},\ \bibinfo {pages} {062002} (\bibinfo {year} {2009})}\BibitemShut
  {NoStop}%
\bibitem [{\citenamefont {Tinto}\ and\ \citenamefont {Dhurandhar}(2005)}]{tdi}%
  \BibitemOpen
  \bibfield  {author} {\bibinfo {author} {\bibfnamefont {M.}~\bibnamefont
  {Tinto}}\ and\ \bibinfo {author} {\bibfnamefont {S.~V.}\ \bibnamefont
  {Dhurandhar}},\ }\href {\doibase 10.12942/lrr-2005-4} {\bibfield  {journal}
  {\bibinfo  {journal} {Living Reviews in Relativity}\ }\textbf {\bibinfo
  {volume} {8}},\ \bibinfo {pages} {4} (\bibinfo {year} {2005})}\BibitemShut
  {NoStop}%
\bibitem [{\citenamefont {Prince}\ \emph {et~al.}(2002)\citenamefont {Prince},
  \citenamefont {Tinto}, \citenamefont {Larson},\ and\ \citenamefont
  {Armstrong}}]{aet}%
  \BibitemOpen
  \bibfield  {author} {\bibinfo {author} {\bibfnamefont {T.~A.}\ \bibnamefont
  {Prince}}, \bibinfo {author} {\bibfnamefont {M.}~\bibnamefont {Tinto}},
  \bibinfo {author} {\bibfnamefont {S.~L.}\ \bibnamefont {Larson}}, \ and\
  \bibinfo {author} {\bibfnamefont {J.~W.}\ \bibnamefont {Armstrong}},\ }\href
  {\doibase 10.1103/PhysRevD.66.122002} {\bibfield  {journal} {\bibinfo
  {journal} {Phys. Rev. D}\ }\textbf {\bibinfo {volume} {66}},\ \bibinfo
  {pages} {122002} (\bibinfo {year} {2002})}\BibitemShut {NoStop}%
\bibitem [{\citenamefont {Armano}\ \emph {et~al.}(2016)\citenamefont {Armano}
  \emph {et~al.}}]{lpf_prl}%
  \BibitemOpen
  \bibfield  {author} {\bibinfo {author} {\bibfnamefont {M.}~\bibnamefont
  {Armano}} \emph {et~al.},\ }\href {\doibase 10.1103/PhysRevLett.116.231101}
  {\bibfield  {journal} {\bibinfo  {journal} {Phys. Rev. Lett.}\ }\textbf
  {\bibinfo {volume} {116}},\ \bibinfo {pages} {231101} (\bibinfo {year}
  {2016})}\BibitemShut {NoStop}%
\bibitem [{\citenamefont {Robson}\ and\ \citenamefont
  {Cornish}(2017)}]{Robson_2017}%
  \BibitemOpen
  \bibfield  {author} {\bibinfo {author} {\bibfnamefont {T.}~\bibnamefont
  {Robson}}\ and\ \bibinfo {author} {\bibfnamefont {N.}~\bibnamefont
  {Cornish}},\ }\href {\doibase 10.1088/1361-6382/aa9601} {\bibfield  {journal}
  {\bibinfo  {journal} {Classical and Quantum Gravity}\ }\textbf {\bibinfo
  {volume} {34}},\ \bibinfo {pages} {244002} (\bibinfo {year}
  {2017})}\BibitemShut {NoStop}%
\bibitem [{\citenamefont {Armano}\ \emph
  {et~al.}(2018{\natexlab{a}})\citenamefont {Armano} \emph
  {et~al.}}]{lpf_prl2}%
  \BibitemOpen
  \bibfield  {author} {\bibinfo {author} {\bibfnamefont {M.}~\bibnamefont
  {Armano}} \emph {et~al.},\ }\href {\doibase 10.1103/PhysRevLett.120.061101}
  {\bibfield  {journal} {\bibinfo  {journal} {Phys. Rev. Lett.}\ }\textbf
  {\bibinfo {volume} {120}},\ \bibinfo {pages} {061101} (\bibinfo {year}
  {2018}{\natexlab{a}})}\BibitemShut {NoStop}%
\bibitem [{\citenamefont {Armano}\ \emph
  {et~al.}(2018{\natexlab{b}})\citenamefont {Armano} \emph {et~al.}}]{lpf_cal}%
  \BibitemOpen
  \bibfield  {author} {\bibinfo {author} {\bibfnamefont {M.}~\bibnamefont
  {Armano}} \emph {et~al.},\ }\href {\doibase 10.1103/PhysRevD.97.122002}
  {\bibfield  {journal} {\bibinfo  {journal} {Phys. Rev. D}\ }\textbf {\bibinfo
  {volume} {97}},\ \bibinfo {pages} {122002} (\bibinfo {year}
  {2018}{\natexlab{b}})}\BibitemShut {NoStop}%
\bibitem [{\citenamefont {Baghi}\ \emph {et~al.}(2022)\citenamefont {Baghi},
  \citenamefont {Korsakova}, \citenamefont {Slutsky}, \citenamefont {Castelli},
  \citenamefont {Karnesis},\ and\ \citenamefont {Bayle}}]{glitches}%
  \BibitemOpen
  \bibfield  {author} {\bibinfo {author} {\bibfnamefont {Q.}~\bibnamefont
  {Baghi}}, \bibinfo {author} {\bibfnamefont {N.}~\bibnamefont {Korsakova}},
  \bibinfo {author} {\bibfnamefont {J.}~\bibnamefont {Slutsky}}, \bibinfo
  {author} {\bibfnamefont {E.}~\bibnamefont {Castelli}}, \bibinfo {author}
  {\bibfnamefont {N.}~\bibnamefont {Karnesis}}, \ and\ \bibinfo {author}
  {\bibfnamefont {J.-B.}\ \bibnamefont {Bayle}},\ }\href {\doibase
  10.1103/PhysRevD.105.042002} {\bibfield  {journal} {\bibinfo  {journal}
  {Phys. Rev. D}\ }\textbf {\bibinfo {volume} {105}},\ \bibinfo {pages}
  {042002} (\bibinfo {year} {2022})}\BibitemShut {NoStop}%
\bibitem [{\citenamefont {Nofrarias}\ \emph {et~al.}(2016)\citenamefont
  {Nofrarias}, \citenamefont {Karnesis} \emph {et~al.}}]{optimaldes}%
  \BibitemOpen
  \bibfield  {author} {\bibinfo {author} {\bibfnamefont {M.}~\bibnamefont
  {Nofrarias}}, \bibinfo {author} {\bibfnamefont {N.}~\bibnamefont {Karnesis}},
   \emph {et~al.},\ }\href {\doibase 10.1103/PhysRevD.93.102004} {\bibfield
  {journal} {\bibinfo  {journal} {Phys. Rev. D}\ }\textbf {\bibinfo {volume}
  {93}},\ \bibinfo {pages} {102004} (\bibinfo {year} {2016})}\BibitemShut
  {NoStop}%
\bibitem [{\citenamefont {Karnesis}\ \emph {et~al.}(2023)\citenamefont
  {Karnesis}, \citenamefont {Katz}, \citenamefont {Korsakova}, \citenamefont
  {Gair},\ and\ \citenamefont {Stergioulas}}]{Karnesis2023ras}%
  \BibitemOpen
  \bibfield  {author} {\bibinfo {author} {\bibfnamefont {N.}~\bibnamefont
  {Karnesis}}, \bibinfo {author} {\bibfnamefont {M.~L.}\ \bibnamefont {Katz}},
  \bibinfo {author} {\bibfnamefont {N.}~\bibnamefont {Korsakova}}, \bibinfo
  {author} {\bibfnamefont {J.~R.}\ \bibnamefont {Gair}}, \ and\ \bibinfo
  {author} {\bibfnamefont {N.}~\bibnamefont {Stergioulas}},\ }\href@noop {} {\
  (\bibinfo {year} {2023})},\ \Eprint {http://arxiv.org/abs/2303.02164}
  {arXiv:2303.02164 [astro-ph.IM]} \BibitemShut {NoStop}%
\bibitem [{\citenamefont {Dey}\ \emph {et~al.}(2021)\citenamefont {Dey},
  \citenamefont {Karnesis} \emph {et~al.}}]{auth5_gaps}%
  \BibitemOpen
  \bibfield  {author} {\bibinfo {author} {\bibfnamefont {K.}~\bibnamefont
  {Dey}}, \bibinfo {author} {\bibfnamefont {N.}~\bibnamefont {Karnesis}},
  \emph {et~al.},\ }\href {\doibase 10.1103/PhysRevD.104.044035} {\bibfield
  {journal} {\bibinfo  {journal} {Phys. Rev. D}\ }\textbf {\bibinfo {volume}
  {104}},\ \bibinfo {pages} {044035} (\bibinfo {year} {2021})}\BibitemShut
  {NoStop}%
\bibitem [{\citenamefont {Baghi}\ \emph {et~al.}(2019)\citenamefont {Baghi},
  \citenamefont {Thorpe}, \citenamefont {Slutsky}, \citenamefont {Baker},
  \citenamefont {Canton}, \citenamefont {Korsakova},\ and\ \citenamefont
  {Karnesis}}]{auth9_gap2}%
  \BibitemOpen
  \bibfield  {author} {\bibinfo {author} {\bibfnamefont {Q.}~\bibnamefont
  {Baghi}}, \bibinfo {author} {\bibfnamefont {J.~I.}\ \bibnamefont {Thorpe}},
  \bibinfo {author} {\bibfnamefont {J.}~\bibnamefont {Slutsky}}, \bibinfo
  {author} {\bibfnamefont {J.}~\bibnamefont {Baker}}, \bibinfo {author}
  {\bibfnamefont {T.~D.}\ \bibnamefont {Canton}}, \bibinfo {author}
  {\bibfnamefont {N.}~\bibnamefont {Korsakova}}, \ and\ \bibinfo {author}
  {\bibfnamefont {N.}~\bibnamefont {Karnesis}},\ }\href {\doibase
  10.1103/PhysRevD.100.022003} {\bibfield  {journal} {\bibinfo  {journal}
  {Phys. Rev. D}\ }\textbf {\bibinfo {volume} {100}},\ \bibinfo {pages}
  {022003} (\bibinfo {year} {2019})}\BibitemShut {NoStop}%
\bibitem [{\citenamefont {Sasli}\ \emph {et~al.}(2023)\citenamefont {Sasli},
  \citenamefont {Karnesis},\ and\ \citenamefont {Stergioulas}}]{Sasli2023mxr}%
  \BibitemOpen
  \bibfield  {author} {\bibinfo {author} {\bibfnamefont {A.}~\bibnamefont
  {Sasli}}, \bibinfo {author} {\bibfnamefont {N.}~\bibnamefont {Karnesis}}, \
  and\ \bibinfo {author} {\bibfnamefont {N.}~\bibnamefont {Stergioulas}},\
  }\href@noop {} {\  (\bibinfo {year} {2023})},\ \Eprint
  {http://arxiv.org/abs/2305.04709} {arXiv:2305.04709 [gr-qc]} \BibitemShut
  {NoStop}%
\bibitem [{\citenamefont {Seoane}\ \emph {et~al.}(2022)\citenamefont {Seoane}
  \emph {et~al.}}]{auth4_missdur}%
  \BibitemOpen
  \bibfield  {author} {\bibinfo {author} {\bibfnamefont {P.~A.}\ \bibnamefont
  {Seoane}} \emph {et~al.},\ }\href {\doibase 10.1007/s10714-021-02889-x}
  {\bibfield  {journal} {\bibinfo  {journal} {Gen. Rel. Grav.}\ }\textbf
  {\bibinfo {volume} {54}},\ \bibinfo {pages} {3} (\bibinfo {year} {2022})},\
  \Eprint {http://arxiv.org/abs/2107.09665} {arXiv:2107.09665 [astro-ph.IM]}
  \BibitemShut {NoStop}%
\bibitem [{\citenamefont {{Sotiriou}}\ and\ \citenamefont
  {{Apostolatos}}(2005)}]{Sotiriou05}%
  \BibitemOpen
  \bibfield  {author} {\bibinfo {author} {\bibfnamefont {T.~P.}\ \bibnamefont
  {{Sotiriou}}}\ and\ \bibinfo {author} {\bibfnamefont {T.~A.}\ \bibnamefont
  {{Apostolatos}}},\ }\href {\doibase 10.1103/PhysRevD.71.044005} {\bibfield
  {journal} {\bibinfo  {journal} {\prd}\ }\textbf {\bibinfo {volume} {71}},\
  \bibinfo {eid} {044005} (\bibinfo {year} {2005})},\ \Eprint
  {http://arxiv.org/abs/gr-qc/0410102} {arXiv:gr-qc/0410102 [gr-qc]}
  \BibitemShut {NoStop}%
\bibitem [{\citenamefont {{Glampedakis}}\ and\ \citenamefont
  {{Apostolatos}}(2013)}]{Glampedakis13}%
  \BibitemOpen
  \bibfield  {author} {\bibinfo {author} {\bibfnamefont {K.}~\bibnamefont
  {{Glampedakis}}}\ and\ \bibinfo {author} {\bibfnamefont {T.~A.}\ \bibnamefont
  {{Apostolatos}}},\ }\href {\doibase 10.1088/0264-9381/30/5/055006} {\bibfield
   {journal} {\bibinfo  {journal} {Classical and Quantum Gravity}\ }\textbf
  {\bibinfo {volume} {30}},\ \bibinfo {eid} {055006} (\bibinfo {year}
  {2013})},\ \Eprint {http://arxiv.org/abs/1301.3309} {arXiv:1301.3309 [gr-qc]}
  \BibitemShut {NoStop}%
\bibitem [{\citenamefont {{Eleni}}\ and\ \citenamefont
  {{Apostolatos}}(2020)}]{Eleni20}%
  \BibitemOpen
  \bibfield  {author} {\bibinfo {author} {\bibfnamefont {A.}~\bibnamefont
  {{Eleni}}}\ and\ \bibinfo {author} {\bibfnamefont {T.~A.}\ \bibnamefont
  {{Apostolatos}}},\ }\href {\doibase 10.1103/PhysRevD.101.044056} {\bibfield
  {journal} {\bibinfo  {journal} {\prd}\ }\textbf {\bibinfo {volume} {101}},\
  \bibinfo {eid} {044056} (\bibinfo {year} {2020})}\BibitemShut {NoStop}%
\bibitem [{\citenamefont {Destounis}\ \emph {et~al.}(2020)\citenamefont
  {Destounis}, \citenamefont {Suvorov},\ and\ \citenamefont
  {Kokkotas}}]{Destounis.2020kss}%
  \BibitemOpen
  \bibfield  {author} {\bibinfo {author} {\bibfnamefont {K.}~\bibnamefont
  {Destounis}}, \bibinfo {author} {\bibfnamefont {A.~G.}\ \bibnamefont
  {Suvorov}}, \ and\ \bibinfo {author} {\bibfnamefont {K.~D.}\ \bibnamefont
  {Kokkotas}},\ }\href {\doibase 10.1103/PhysRevD.102.064041} {\bibfield
  {journal} {\bibinfo  {journal} {Phys. Rev. D}\ }\textbf {\bibinfo {volume}
  {102}},\ \bibinfo {pages} {064041} (\bibinfo {year} {2020})},\ \Eprint
  {http://arxiv.org/abs/2009.00028} {arXiv:2009.00028 [gr-qc]} \BibitemShut
  {NoStop}%
\bibitem [{\citenamefont {{Timogiannis}}\ \emph {et~al.}(2021)\citenamefont
  {{Timogiannis}}, \citenamefont {{Lukes-Gerakopoulos}},\ and\ \citenamefont
  {{Apostolatos}}}]{Timogiannis21}%
  \BibitemOpen
  \bibfield  {author} {\bibinfo {author} {\bibfnamefont {I.}~\bibnamefont
  {{Timogiannis}}}, \bibinfo {author} {\bibfnamefont {G.}~\bibnamefont
  {{Lukes-Gerakopoulos}}}, \ and\ \bibinfo {author} {\bibfnamefont {T.~A.}\
  \bibnamefont {{Apostolatos}}},\ }\href {\doibase 10.1103/PhysRevD.104.024042}
  {\bibfield  {journal} {\bibinfo  {journal} {\prd}\ }\textbf {\bibinfo
  {volume} {104}},\ \bibinfo {eid} {024042} (\bibinfo {year} {2021})},\ \Eprint
  {http://arxiv.org/abs/2104.11183} {arXiv:2104.11183 [gr-qc]} \BibitemShut
  {NoStop}%
\bibitem [{\citenamefont {{Apostolatos}}\ \emph {et~al.}(2009)\citenamefont
  {{Apostolatos}}, \citenamefont {{Lukes-Gerakopoulos}},\ and\ \citenamefont
  {{Contopoulos}}}]{Apostolatos09}%
  \BibitemOpen
  \bibfield  {author} {\bibinfo {author} {\bibfnamefont {T.~A.}\ \bibnamefont
  {{Apostolatos}}}, \bibinfo {author} {\bibfnamefont {G.}~\bibnamefont
  {{Lukes-Gerakopoulos}}}, \ and\ \bibinfo {author} {\bibfnamefont
  {G.}~\bibnamefont {{Contopoulos}}},\ }\href {\doibase
  10.1103/PhysRevLett.103.111101} {\bibfield  {journal} {\bibinfo  {journal}
  {\prl}\ }\textbf {\bibinfo {volume} {103}},\ \bibinfo {eid} {111101}
  (\bibinfo {year} {2009})},\ \Eprint {http://arxiv.org/abs/0906.0093}
  {arXiv:0906.0093 [gr-qc]} \BibitemShut {NoStop}%
\bibitem [{\citenamefont {Destounis}\ \emph {et~al.}(2021)\citenamefont
  {Destounis}, \citenamefont {Suvorov},\ and\ \citenamefont
  {Kokkotas}}]{Destounis.2021mqv}%
  \BibitemOpen
  \bibfield  {author} {\bibinfo {author} {\bibfnamefont {K.}~\bibnamefont
  {Destounis}}, \bibinfo {author} {\bibfnamefont {A.~G.}\ \bibnamefont
  {Suvorov}}, \ and\ \bibinfo {author} {\bibfnamefont {K.~D.}\ \bibnamefont
  {Kokkotas}},\ }\href {\doibase 10.1103/PhysRevLett.126.141102} {\bibfield
  {journal} {\bibinfo  {journal} {Phys. Rev. Lett.}\ }\textbf {\bibinfo
  {volume} {126}},\ \bibinfo {pages} {141102} (\bibinfo {year} {2021})},\
  \Eprint {http://arxiv.org/abs/2103.05643} {arXiv:2103.05643 [gr-qc]}
  \BibitemShut {NoStop}%
\bibitem [{\citenamefont {Destounis}\ and\ \citenamefont
  {Kokkotas}(2021)}]{Destounis.2021rko}%
  \BibitemOpen
  \bibfield  {author} {\bibinfo {author} {\bibfnamefont {K.}~\bibnamefont
  {Destounis}}\ and\ \bibinfo {author} {\bibfnamefont {K.~D.}\ \bibnamefont
  {Kokkotas}},\ }\href {\doibase 10.1103/PhysRevD.104.064023} {\bibfield
  {journal} {\bibinfo  {journal} {Phys. Rev. D}\ }\textbf {\bibinfo {volume}
  {104}},\ \bibinfo {pages} {064023} (\bibinfo {year} {2021})},\ \Eprint
  {http://arxiv.org/abs/2108.02782} {arXiv:2108.02782 [gr-qc]} \BibitemShut
  {NoStop}%
\bibitem [{\citenamefont {{Lyutikov}}(2005)}]{2005MNRAS.362.1078L}%
  \BibitemOpen
  \bibfield  {author} {\bibinfo {author} {\bibfnamefont {M.}~\bibnamefont
  {{Lyutikov}}},\ }\href {\doibase 10.1111/j.1365-2966.2005.09388.x} {\bibfield
   {journal} {\bibinfo  {journal} {\mnras}\ }\textbf {\bibinfo {volume}
  {362}},\ \bibinfo {pages} {1078} (\bibinfo {year} {2005})},\ \Eprint
  {http://arxiv.org/abs/astro-ph/0502503} {arXiv:astro-ph/0502503 [astro-ph]}
  \BibitemShut {NoStop}%
\bibitem [{\citenamefont {{Perera}}\ \emph {et~al.}(2012)\citenamefont
  {{Perera}}, \citenamefont {{Lomiashvili}}, \citenamefont {{Gourgouliatos}},
  \citenamefont {{McLaughlin}},\ and\ \citenamefont
  {{Lyutikov}}}]{2012ApJ...750..130P}%
  \BibitemOpen
  \bibfield  {author} {\bibinfo {author} {\bibfnamefont {B.~B.~P.}\
  \bibnamefont {{Perera}}}, \bibinfo {author} {\bibfnamefont {D.}~\bibnamefont
  {{Lomiashvili}}}, \bibinfo {author} {\bibfnamefont {K.~N.}\ \bibnamefont
  {{Gourgouliatos}}}, \bibinfo {author} {\bibfnamefont {M.~A.}\ \bibnamefont
  {{McLaughlin}}}, \ and\ \bibinfo {author} {\bibfnamefont {M.}~\bibnamefont
  {{Lyutikov}}},\ }\href {\doibase 10.1088/0004-637X/750/2/130} {\bibfield
  {journal} {\bibinfo  {journal} {\apj}\ }\textbf {\bibinfo {volume} {750}},\
  \bibinfo {eid} {130} (\bibinfo {year} {2012})},\ \Eprint
  {http://arxiv.org/abs/1203.0763} {arXiv:1203.0763 [astro-ph.GA]} \BibitemShut
  {NoStop}%
\bibitem [{\citenamefont {{Tsang}}(2013)}]{2013ApJ...777..103T}%
  \BibitemOpen
  \bibfield  {author} {\bibinfo {author} {\bibfnamefont {D.}~\bibnamefont
  {{Tsang}}},\ }\href {\doibase 10.1088/0004-637X/777/2/103} {\bibfield
  {journal} {\bibinfo  {journal} {\apj}\ }\textbf {\bibinfo {volume} {777}},\
  \bibinfo {eid} {103} (\bibinfo {year} {2013})},\ \Eprint
  {http://arxiv.org/abs/1307.3554} {arXiv:1307.3554 [astro-ph.HE]} \BibitemShut
  {NoStop}%
\bibitem [{\citenamefont {{Gourgouliatos}}\ and\ \citenamefont
  {{Lynden-Bell}}(2019)}]{2019MNRAS.482.1942G}%
  \BibitemOpen
  \bibfield  {author} {\bibinfo {author} {\bibfnamefont {K.~N.}\ \bibnamefont
  {{Gourgouliatos}}}\ and\ \bibinfo {author} {\bibfnamefont {D.}~\bibnamefont
  {{Lynden-Bell}}},\ }\href {\doibase 10.1093/mnras/sty2766} {\bibfield
  {journal} {\bibinfo  {journal} {\mnras}\ }\textbf {\bibinfo {volume} {482}},\
  \bibinfo {pages} {1942} (\bibinfo {year} {2019})},\ \Eprint
  {http://arxiv.org/abs/1810.04177} {arXiv:1810.04177 [astro-ph.HE]}
  \BibitemShut {NoStop}%
\bibitem [{\citenamefont {{Sur}}\ and\ \citenamefont
  {{Haskell}}(2021)}]{2021MNRAS.502.4680S}%
  \BibitemOpen
  \bibfield  {author} {\bibinfo {author} {\bibfnamefont {A.}~\bibnamefont
  {{Sur}}}\ and\ \bibinfo {author} {\bibfnamefont {B.}~\bibnamefont
  {{Haskell}}},\ }\href {\doibase 10.1093/mnras/stab307} {\bibfield  {journal}
  {\bibinfo  {journal} {\mnras}\ }\textbf {\bibinfo {volume} {502}},\ \bibinfo
  {pages} {4680} (\bibinfo {year} {2021})},\ \Eprint
  {http://arxiv.org/abs/2010.15574} {arXiv:2010.15574 [astro-ph.HE]}
  \BibitemShut {NoStop}%
\bibitem [{\citenamefont {{Gourgouliatos}}\ and\ \citenamefont
  {{Hollerbach}}(2018)}]{2018ApJ...852...21G}%
  \BibitemOpen
  \bibfield  {author} {\bibinfo {author} {\bibfnamefont {K.~N.}\ \bibnamefont
  {{Gourgouliatos}}}\ and\ \bibinfo {author} {\bibfnamefont {R.}~\bibnamefont
  {{Hollerbach}}},\ }\href {\doibase 10.3847/1538-4357/aa9d93} {\bibfield
  {journal} {\bibinfo  {journal} {\apj}\ }\textbf {\bibinfo {volume} {852}},\
  \bibinfo {eid} {21} (\bibinfo {year} {2018})},\ \Eprint
  {http://arxiv.org/abs/1710.01338} {arXiv:1710.01338 [astro-ph.HE]}
  \BibitemShut {NoStop}%
\bibitem [{\citenamefont {{Gourgouliatos}}\ \emph {et~al.}(2020)\citenamefont
  {{Gourgouliatos}}, \citenamefont {{Hollerbach}},\ and\ \citenamefont
  {{Igoshev}}}]{2020MNRAS.495.1692G}%
  \BibitemOpen
  \bibfield  {author} {\bibinfo {author} {\bibfnamefont {K.~N.}\ \bibnamefont
  {{Gourgouliatos}}}, \bibinfo {author} {\bibfnamefont {R.}~\bibnamefont
  {{Hollerbach}}}, \ and\ \bibinfo {author} {\bibfnamefont {A.~P.}\
  \bibnamefont {{Igoshev}}},\ }\href {\doibase 10.1093/mnras/staa1295}
  {\bibfield  {journal} {\bibinfo  {journal} {\mnras}\ }\textbf {\bibinfo
  {volume} {495}},\ \bibinfo {pages} {1692} (\bibinfo {year} {2020})},\ \Eprint
  {http://arxiv.org/abs/2005.02410} {arXiv:2005.02410 [astro-ph.HE]}
  \BibitemShut {NoStop}%
\bibitem [{\citenamefont {{Igoshev}}\ \emph {et~al.}(2021)\citenamefont
  {{Igoshev}}, \citenamefont {{Gourgouliatos}}, \citenamefont {{Hollerbach}},\
  and\ \citenamefont {{Wood}}}]{2021ApJ...909..101I}%
  \BibitemOpen
  \bibfield  {author} {\bibinfo {author} {\bibfnamefont {A.~P.}\ \bibnamefont
  {{Igoshev}}}, \bibinfo {author} {\bibfnamefont {K.~N.}\ \bibnamefont
  {{Gourgouliatos}}}, \bibinfo {author} {\bibfnamefont {R.}~\bibnamefont
  {{Hollerbach}}}, \ and\ \bibinfo {author} {\bibfnamefont {T.~S.}\
  \bibnamefont {{Wood}}},\ }\href {\doibase 10.3847/1538-4357/abde3e}
  {\bibfield  {journal} {\bibinfo  {journal} {\apj}\ }\textbf {\bibinfo
  {volume} {909}},\ \bibinfo {eid} {101} (\bibinfo {year} {2021})},\ \Eprint
  {http://arxiv.org/abs/2101.08292} {arXiv:2101.08292 [astro-ph.HE]}
  \BibitemShut {NoStop}%
\bibitem [{\citenamefont {{Benacquista}}\ and\ \citenamefont
  {{Holley-Bockelmann}}(2006)}]{gal1}%
  \BibitemOpen
  \bibfield  {author} {\bibinfo {author} {\bibfnamefont {M.}~\bibnamefont
  {{Benacquista}}}\ and\ \bibinfo {author} {\bibfnamefont {K.}~\bibnamefont
  {{Holley-Bockelmann}}},\ }\href {\doibase 10.1086/504024} {\bibfield
  {journal} {\bibinfo  {journal} {\apj}\ }\textbf {\bibinfo {volume} {645}},\
  \bibinfo {pages} {589} (\bibinfo {year} {2006})},\ \Eprint
  {http://arxiv.org/abs/astro-ph/0504135} {arXiv:astro-ph/0504135 [astro-ph]}
  \BibitemShut {NoStop}%
\bibitem [{\citenamefont {{Korol}}\ \emph {et~al.}(2022)\citenamefont
  {{Korol}}, \citenamefont {{Hallakoun}}, \citenamefont {{Toonen}},\ and\
  \citenamefont {{Karnesis}}}]{gal2}%
  \BibitemOpen
  \bibfield  {author} {\bibinfo {author} {\bibfnamefont {V.}~\bibnamefont
  {{Korol}}}, \bibinfo {author} {\bibfnamefont {N.}~\bibnamefont
  {{Hallakoun}}}, \bibinfo {author} {\bibfnamefont {S.}~\bibnamefont
  {{Toonen}}}, \ and\ \bibinfo {author} {\bibfnamefont {N.}~\bibnamefont
  {{Karnesis}}},\ }\href {\doibase 10.1093/mnras/stac415} {\bibfield  {journal}
  {\bibinfo  {journal} {\mnras}\ }\textbf {\bibinfo {volume} {511}},\ \bibinfo
  {pages} {5936} (\bibinfo {year} {2022})},\ \Eprint
  {http://arxiv.org/abs/2109.10972} {arXiv:2109.10972 [astro-ph.HE]}
  \BibitemShut {NoStop}%
\bibitem [{\citenamefont {{Korol}}\ \emph {et~al.}(2017)\citenamefont
  {{Korol}}, \citenamefont {{Rossi}}, \citenamefont {{Groot}}, \citenamefont
  {{Nelemans}}, \citenamefont {{Toonen}},\ and\ \citenamefont
  {{Brown}}}]{gal3}%
  \BibitemOpen
  \bibfield  {author} {\bibinfo {author} {\bibfnamefont {V.}~\bibnamefont
  {{Korol}}}, \bibinfo {author} {\bibfnamefont {E.~M.}\ \bibnamefont
  {{Rossi}}}, \bibinfo {author} {\bibfnamefont {P.~J.}\ \bibnamefont
  {{Groot}}}, \bibinfo {author} {\bibfnamefont {G.}~\bibnamefont {{Nelemans}}},
  \bibinfo {author} {\bibfnamefont {S.}~\bibnamefont {{Toonen}}}, \ and\
  \bibinfo {author} {\bibfnamefont {A.~G.~A.}\ \bibnamefont {{Brown}}},\ }\href
  {\doibase 10.1093/mnras/stx1285} {\bibfield  {journal} {\bibinfo  {journal}
  {\mnras}\ }\textbf {\bibinfo {volume} {470}},\ \bibinfo {pages} {1894}
  (\bibinfo {year} {2017})},\ \Eprint {http://arxiv.org/abs/1703.02555}
  {arXiv:1703.02555 [astro-ph.HE]} \BibitemShut {NoStop}%
\bibitem [{\citenamefont {Georgousi}\ \emph {et~al.}(2022)\citenamefont
  {Georgousi}, \citenamefont {Karnesis}, \citenamefont {Korol}, \citenamefont
  {Pieroni},\ and\ \citenamefont {Stergioulas}}]{auth1_geor}%
  \BibitemOpen
  \bibfield  {author} {\bibinfo {author} {\bibfnamefont {M.}~\bibnamefont
  {Georgousi}}, \bibinfo {author} {\bibfnamefont {N.}~\bibnamefont {Karnesis}},
  \bibinfo {author} {\bibfnamefont {V.}~\bibnamefont {Korol}}, \bibinfo
  {author} {\bibfnamefont {M.}~\bibnamefont {Pieroni}}, \ and\ \bibinfo
  {author} {\bibfnamefont {N.}~\bibnamefont {Stergioulas}},\ }\href {\doibase
  10.1093/mnras/stac3686} {\bibfield  {journal} {\bibinfo  {journal} {Mon. Not.
  Roy. Astron. Soc.}\ }\textbf {\bibinfo {volume} {519}},\ \bibinfo {pages}
  {2552} (\bibinfo {year} {2022})},\ \Eprint {http://arxiv.org/abs/2204.07349}
  {arXiv:2204.07349 [astro-ph.GA]} \BibitemShut {NoStop}%
\bibitem [{\citenamefont {Katz}\ \emph {et~al.}(2022)\citenamefont {Katz},
  \citenamefont {Danielski}, \citenamefont {Karnesis}, \citenamefont {Korol},
  \citenamefont {Tamanini}, \citenamefont {Cornish},\ and\ \citenamefont
  {Littenberg}}]{auth2_triplets}%
  \BibitemOpen
  \bibfield  {author} {\bibinfo {author} {\bibfnamefont {M.~L.}\ \bibnamefont
  {Katz}}, \bibinfo {author} {\bibfnamefont {C.}~\bibnamefont {Danielski}},
  \bibinfo {author} {\bibfnamefont {N.}~\bibnamefont {Karnesis}}, \bibinfo
  {author} {\bibfnamefont {V.}~\bibnamefont {Korol}}, \bibinfo {author}
  {\bibfnamefont {N.}~\bibnamefont {Tamanini}}, \bibinfo {author}
  {\bibfnamefont {N.~J.}\ \bibnamefont {Cornish}}, \ and\ \bibinfo {author}
  {\bibfnamefont {T.~B.}\ \bibnamefont {Littenberg}},\ }\href@noop {} {\
  (\bibinfo {year} {2022})},\ \Eprint {http://arxiv.org/abs/2205.03461}
  {arXiv:2205.03461 [astro-ph.EP]} \BibitemShut {NoStop}%
\bibitem [{\citenamefont {Korol}\ \emph {et~al.}(2022)\citenamefont {Korol},
  \citenamefont {Hallakoun}, \citenamefont {Toonen},\ and\ \citenamefont
  {Karnesis}}]{auth3_obs}%
  \BibitemOpen
  \bibfield  {author} {\bibinfo {author} {\bibfnamefont {V.}~\bibnamefont
  {Korol}}, \bibinfo {author} {\bibfnamefont {N.}~\bibnamefont {Hallakoun}},
  \bibinfo {author} {\bibfnamefont {S.}~\bibnamefont {Toonen}}, \ and\ \bibinfo
  {author} {\bibfnamefont {N.}~\bibnamefont {Karnesis}},\ }\href {\doibase
  10.1093/mnras/stac415} {\bibfield  {journal} {\bibinfo  {journal} {Monthly
  Notices of the Royal Astronomical Society}\ }\textbf {\bibinfo {volume}
  {511}},\ \bibinfo {pages} {5936} (\bibinfo {year} {2022})},\ \Eprint
  {http://arxiv.org/abs/https://academic.oup.com/mnras/article-pdf/511/4/5936/42794528/stac415.pdf}
  {https://academic.oup.com/mnras/article-pdf/511/4/5936/42794528/stac415.pdf}
  \BibitemShut {NoStop}%
\bibitem [{\citenamefont {Littenberg}\ and\ \citenamefont
  {Yunes}(2019)}]{gal4}%
  \BibitemOpen
  \bibfield  {author} {\bibinfo {author} {\bibfnamefont {T.~B.}\ \bibnamefont
  {Littenberg}}\ and\ \bibinfo {author} {\bibfnamefont {N.}~\bibnamefont
  {Yunes}},\ }\href {\doibase 10.1088/1361-6382/ab0a3d} {\bibfield  {journal}
  {\bibinfo  {journal} {Class. Quant. Grav.}\ }\textbf {\bibinfo {volume}
  {36}},\ \bibinfo {pages} {095017} (\bibinfo {year} {2019})},\ \Eprint
  {http://arxiv.org/abs/1811.01093} {arXiv:1811.01093 [gr-qc]} \BibitemShut
  {NoStop}%
\bibitem [{\citenamefont {Karnesis}\ \emph {et~al.}(2021)\citenamefont
  {Karnesis}, \citenamefont {Babak}, \citenamefont {Pieroni}, \citenamefont
  {Cornish},\ and\ \citenamefont {Littenberg}}]{auth6_gal}%
  \BibitemOpen
  \bibfield  {author} {\bibinfo {author} {\bibfnamefont {N.}~\bibnamefont
  {Karnesis}}, \bibinfo {author} {\bibfnamefont {S.}~\bibnamefont {Babak}},
  \bibinfo {author} {\bibfnamefont {M.}~\bibnamefont {Pieroni}}, \bibinfo
  {author} {\bibfnamefont {N.}~\bibnamefont {Cornish}}, \ and\ \bibinfo
  {author} {\bibfnamefont {T.}~\bibnamefont {Littenberg}},\ }\href {\doibase
  10.1103/PhysRevD.104.043019} {\bibfield  {journal} {\bibinfo  {journal}
  {Phys. Rev. D}\ }\textbf {\bibinfo {volume} {104}},\ \bibinfo {pages}
  {043019} (\bibinfo {year} {2021})}\BibitemShut {NoStop}%
\bibitem [{\citenamefont {Oikonomou}(2023)}]{oikonomou1}%
  \BibitemOpen
  \bibfield  {author} {\bibinfo {author} {\bibfnamefont {V.~K.}\ \bibnamefont
  {Oikonomou}},\ }\href {\doibase 10.1016/j.astropartphys.2022.102777}
  {\bibfield  {journal} {\bibinfo  {journal} {Astropart. Phys.}\ }\textbf
  {\bibinfo {volume} {144}},\ \bibinfo {pages} {102777} (\bibinfo {year}
  {2023})},\ \Eprint {http://arxiv.org/abs/2209.09781} {arXiv:2209.09781
  [gr-qc]} \BibitemShut {NoStop}%
\bibitem [{\citenamefont {Oikonomou}(2022)}]{oikonomou2}%
  \BibitemOpen
  \bibfield  {author} {\bibinfo {author} {\bibfnamefont {V.~K.}\ \bibnamefont
  {Oikonomou}},\ }\href {\doibase 10.1002/andp.202200134} {\bibfield  {journal}
  {\bibinfo  {journal} {Annalen Phys.}\ }\textbf {\bibinfo {volume} {534}},\
  \bibinfo {pages} {2200134} (\bibinfo {year} {2022})},\ \Eprint
  {http://arxiv.org/abs/2205.15405} {arXiv:2205.15405 [gr-qc]} \BibitemShut
  {NoStop}%
\bibitem [{\citenamefont {Odintsov}\ \emph {et~al.}(2022)\citenamefont
  {Odintsov}, \citenamefont {Oikonomou},\ and\ \citenamefont
  {Myrzakulov}}]{oikonomou3}%
  \BibitemOpen
  \bibfield  {author} {\bibinfo {author} {\bibfnamefont {S.~D.}\ \bibnamefont
  {Odintsov}}, \bibinfo {author} {\bibfnamefont {V.~K.}\ \bibnamefont
  {Oikonomou}}, \ and\ \bibinfo {author} {\bibfnamefont {R.}~\bibnamefont
  {Myrzakulov}},\ }\href {\doibase 10.3390/sym14040729} {\bibfield  {journal}
  {\bibinfo  {journal} {Symmetry}\ }\textbf {\bibinfo {volume} {14}},\ \bibinfo
  {pages} {729} (\bibinfo {year} {2022})},\ \Eprint
  {http://arxiv.org/abs/2204.00876} {arXiv:2204.00876 [gr-qc]} \BibitemShut
  {NoStop}%
\bibitem [{\citenamefont {{Di Valentino}}\ \emph {et~al.}(2021)\citenamefont
  {{Di Valentino}}, \citenamefont {{Mena}}, \citenamefont {{Pan}},
  \citenamefont {{Visinelli}}, \citenamefont {{Yang}}, \citenamefont
  {{Melchiorri}}, \citenamefont {{Mota}}, \citenamefont {{Riess}},\ and\
  \citenamefont {{Silk}}}]{2021CQGra..38o3001D}%
  \BibitemOpen
  \bibfield  {author} {\bibinfo {author} {\bibfnamefont {E.}~\bibnamefont {{Di
  Valentino}}}, \bibinfo {author} {\bibfnamefont {O.}~\bibnamefont {{Mena}}},
  \bibinfo {author} {\bibfnamefont {S.}~\bibnamefont {{Pan}}}, \bibinfo
  {author} {\bibfnamefont {L.}~\bibnamefont {{Visinelli}}}, \bibinfo {author}
  {\bibfnamefont {W.}~\bibnamefont {{Yang}}}, \bibinfo {author} {\bibfnamefont
  {A.}~\bibnamefont {{Melchiorri}}}, \bibinfo {author} {\bibfnamefont {D.~F.}\
  \bibnamefont {{Mota}}}, \bibinfo {author} {\bibfnamefont {A.~G.}\
  \bibnamefont {{Riess}}}, \ and\ \bibinfo {author} {\bibfnamefont
  {J.}~\bibnamefont {{Silk}}},\ }\href {\doibase 10.1088/1361-6382/ac086d}
  {\bibfield  {journal} {\bibinfo  {journal} {Classical and Quantum Gravity}\
  }\textbf {\bibinfo {volume} {38}},\ \bibinfo {eid} {153001} (\bibinfo {year}
  {2021})},\ \Eprint {http://arxiv.org/abs/2103.01183} {arXiv:2103.01183
  [astro-ph.CO]} \BibitemShut {NoStop}%
\bibitem [{\citenamefont {{Perivolaropoulos}}\ and\ \citenamefont
  {{Skara}}(2021{\natexlab{a}})}]{2021arXiv210505208P}%
  \BibitemOpen
  \bibfield  {author} {\bibinfo {author} {\bibfnamefont {L.}~\bibnamefont
  {{Perivolaropoulos}}}\ and\ \bibinfo {author} {\bibfnamefont
  {F.}~\bibnamefont {{Skara}}},\ }\href@noop {} {\bibfield  {journal} {\bibinfo
   {journal} {arXiv e-prints}\ ,\ \bibinfo {eid} {arXiv:2105.05208}} (\bibinfo
  {year} {2021}{\natexlab{a}})},\ \Eprint {http://arxiv.org/abs/2105.05208}
  {arXiv:2105.05208 [astro-ph.CO]} \BibitemShut {NoStop}%
\bibitem [{\citenamefont {Abdalla}\ \emph {et~al.}(2022)\citenamefont {Abdalla}
  \emph {et~al.}}]{2022arXiv220306142A}%
  \BibitemOpen
  \bibfield  {author} {\bibinfo {author} {\bibfnamefont {E.}~\bibnamefont
  {Abdalla}} \emph {et~al.},\ }\href@noop {} {\bibfield  {journal} {\bibinfo
  {journal} {arXiv e-prints}\ ,\ \bibinfo {eid} {arXiv:2203.06142}} (\bibinfo
  {year} {2022})},\ \Eprint {http://arxiv.org/abs/2203.06142} {arXiv:2203.06142
  [astro-ph.CO]} \BibitemShut {NoStop}%
\bibitem [{\citenamefont {{Alestas}}\ and\ \citenamefont
  {{Perivolaropoulos}}(2021)}]{2021MNRAS.504.3956A}%
  \BibitemOpen
  \bibfield  {author} {\bibinfo {author} {\bibfnamefont {G.}~\bibnamefont
  {{Alestas}}}\ and\ \bibinfo {author} {\bibfnamefont {L.}~\bibnamefont
  {{Perivolaropoulos}}},\ }\href {\doibase 10.1093/mnras/stab1070} {\bibfield
  {journal} {\bibinfo  {journal} {Monthly Notices of the Royal Astronomical
  Society}\ }\textbf {\bibinfo {volume} {504}},\ \bibinfo {pages} {3956}
  (\bibinfo {year} {2021})},\ \Eprint {http://arxiv.org/abs/2103.04045}
  {arXiv:2103.04045 [astro-ph.CO]} \BibitemShut {NoStop}%
\bibitem [{\citenamefont {{Marra}}\ and\ \citenamefont
  {{Perivolaropoulos}}(2021)}]{2021PhRvD.104b1303M}%
  \BibitemOpen
  \bibfield  {author} {\bibinfo {author} {\bibfnamefont {V.}~\bibnamefont
  {{Marra}}}\ and\ \bibinfo {author} {\bibfnamefont {L.}~\bibnamefont
  {{Perivolaropoulos}}},\ }\href {\doibase 10.1103/PhysRevD.104.L021303}
  {\bibfield  {journal} {\bibinfo  {journal} {\prd}\ }\textbf {\bibinfo
  {volume} {104}},\ \bibinfo {eid} {L021303} (\bibinfo {year} {2021})},\
  \Eprint {http://arxiv.org/abs/2102.06012} {arXiv:2102.06012 [astro-ph.CO]}
  \BibitemShut {NoStop}%
\bibitem [{\citenamefont {{Alestas}}\ \emph
  {et~al.}(2021{\natexlab{a}})\citenamefont {{Alestas}}, \citenamefont
  {{Kazantzidis}},\ and\ \citenamefont
  {{Perivolaropoulos}}}]{2021PhRvD.103h3517A}%
  \BibitemOpen
  \bibfield  {author} {\bibinfo {author} {\bibfnamefont {G.}~\bibnamefont
  {{Alestas}}}, \bibinfo {author} {\bibfnamefont {L.}~\bibnamefont
  {{Kazantzidis}}}, \ and\ \bibinfo {author} {\bibfnamefont {L.}~\bibnamefont
  {{Perivolaropoulos}}},\ }\href {\doibase 10.1103/PhysRevD.103.083517}
  {\bibfield  {journal} {\bibinfo  {journal} {\prd}\ }\textbf {\bibinfo
  {volume} {103}},\ \bibinfo {eid} {083517} (\bibinfo {year}
  {2021}{\natexlab{a}})},\ \Eprint {http://arxiv.org/abs/2012.13932}
  {arXiv:2012.13932 [astro-ph.CO]} \BibitemShut {NoStop}%
\bibitem [{\citenamefont {{Alestas}}\ \emph
  {et~al.}(2022{\natexlab{a}})\citenamefont {{Alestas}}, \citenamefont
  {{Camarena}}, \citenamefont {{Di Valentino}}, \citenamefont {{Kazantzidis}},
  \citenamefont {{Marra}}, \citenamefont {{Nesseris}},\ and\ \citenamefont
  {{Perivolaropoulos}}}]{2022PhRvD.105f3538A}%
  \BibitemOpen
  \bibfield  {author} {\bibinfo {author} {\bibfnamefont {G.}~\bibnamefont
  {{Alestas}}}, \bibinfo {author} {\bibfnamefont {D.}~\bibnamefont
  {{Camarena}}}, \bibinfo {author} {\bibfnamefont {E.}~\bibnamefont {{Di
  Valentino}}}, \bibinfo {author} {\bibfnamefont {L.}~\bibnamefont
  {{Kazantzidis}}}, \bibinfo {author} {\bibfnamefont {V.}~\bibnamefont
  {{Marra}}}, \bibinfo {author} {\bibfnamefont {S.}~\bibnamefont {{Nesseris}}},
  \ and\ \bibinfo {author} {\bibfnamefont {L.}~\bibnamefont
  {{Perivolaropoulos}}},\ }\href {\doibase 10.1103/PhysRevD.105.063538}
  {\bibfield  {journal} {\bibinfo  {journal} {\prd}\ }\textbf {\bibinfo
  {volume} {105}},\ \bibinfo {eid} {063538} (\bibinfo {year}
  {2022}{\natexlab{a}})},\ \Eprint {http://arxiv.org/abs/2110.04336}
  {arXiv:2110.04336 [astro-ph.CO]} \BibitemShut {NoStop}%
\bibitem [{\citenamefont {{Perivolaropoulos}}\ and\ \citenamefont
  {{Skara}}(2022)}]{2022arXiv220310374P}%
  \BibitemOpen
  \bibfield  {author} {\bibinfo {author} {\bibfnamefont {L.}~\bibnamefont
  {{Perivolaropoulos}}}\ and\ \bibinfo {author} {\bibfnamefont
  {F.}~\bibnamefont {{Skara}}},\ }\href@noop {} {\bibfield  {journal} {\bibinfo
   {journal} {arXiv e-prints}\ ,\ \bibinfo {eid} {arXiv:2203.10374}} (\bibinfo
  {year} {2022})},\ \Eprint {http://arxiv.org/abs/2203.10374} {arXiv:2203.10374
  [astro-ph.CO]} \BibitemShut {NoStop}%
\bibitem [{\citenamefont {{Perivolaropoulos}}\ and\ \citenamefont
  {{Skara}}(2021{\natexlab{b}})}]{2021PhRvD.104l3511P}%
  \BibitemOpen
  \bibfield  {author} {\bibinfo {author} {\bibfnamefont {L.}~\bibnamefont
  {{Perivolaropoulos}}}\ and\ \bibinfo {author} {\bibfnamefont
  {F.}~\bibnamefont {{Skara}}},\ }\href {\doibase 10.1103/PhysRevD.104.123511}
  {\bibfield  {journal} {\bibinfo  {journal} {\prd}\ }\textbf {\bibinfo
  {volume} {104}},\ \bibinfo {eid} {123511} (\bibinfo {year}
  {2021}{\natexlab{b}})},\ \Eprint {http://arxiv.org/abs/2109.04406}
  {arXiv:2109.04406 [astro-ph.CO]} \BibitemShut {NoStop}%
\bibitem [{\citenamefont {{Alestas}}\ \emph
  {et~al.}(2021{\natexlab{b}})\citenamefont {{Alestas}}, \citenamefont
  {{Antoniou}},\ and\ \citenamefont
  {{Perivolaropoulos}}}]{2021Univ....7..366A}%
  \BibitemOpen
  \bibfield  {author} {\bibinfo {author} {\bibfnamefont {G.}~\bibnamefont
  {{Alestas}}}, \bibinfo {author} {\bibfnamefont {I.}~\bibnamefont
  {{Antoniou}}}, \ and\ \bibinfo {author} {\bibfnamefont {L.}~\bibnamefont
  {{Perivolaropoulos}}},\ }\href {\doibase 10.3390/universe7100366} {\bibfield
  {journal} {\bibinfo  {journal} {Universe}\ }\textbf {\bibinfo {volume} {7}},\
  \bibinfo {pages} {366} (\bibinfo {year} {2021}{\natexlab{b}})},\ \Eprint
  {http://arxiv.org/abs/2104.14481} {arXiv:2104.14481 [astro-ph.CO]}
  \BibitemShut {NoStop}%
\bibitem [{\citenamefont {{Alestas}}\ \emph
  {et~al.}(2022{\natexlab{b}})\citenamefont {{Alestas}}, \citenamefont
  {{Perivolaropoulos}},\ and\ \citenamefont {{Tanidis}}}]{2022arXiv220105846A}%
  \BibitemOpen
  \bibfield  {author} {\bibinfo {author} {\bibfnamefont {G.}~\bibnamefont
  {{Alestas}}}, \bibinfo {author} {\bibfnamefont {L.}~\bibnamefont
  {{Perivolaropoulos}}}, \ and\ \bibinfo {author} {\bibfnamefont
  {K.}~\bibnamefont {{Tanidis}}},\ }\href@noop {} {\bibfield  {journal}
  {\bibinfo  {journal} {arXiv e-prints}\ ,\ \bibinfo {eid} {arXiv:2201.05846}}
  (\bibinfo {year} {2022}{\natexlab{b}})},\ \Eprint
  {http://arxiv.org/abs/2201.05846} {arXiv:2201.05846 [astro-ph.CO]}
  \BibitemShut {NoStop}%
\bibitem [{\citenamefont {{Perivolaropoulos}}(2022)}]{2022arXiv220108997P}%
  \BibitemOpen
  \bibfield  {author} {\bibinfo {author} {\bibfnamefont {L.}~\bibnamefont
  {{Perivolaropoulos}}},\ }\href@noop {} {\bibfield  {journal} {\bibinfo
  {journal} {arXiv e-prints}\ ,\ \bibinfo {eid} {arXiv:2201.08997}} (\bibinfo
  {year} {2022})},\ \Eprint {http://arxiv.org/abs/2201.08997} {arXiv:2201.08997
  [astro-ph.EP]} \BibitemShut {NoStop}%
\bibitem [{\citenamefont {Barausse}\ \emph {et~al.}(2020)\citenamefont
  {Barausse} \emph {et~al.}}]{2020GReGr..52...81B}%
  \BibitemOpen
  \bibfield  {author} {\bibinfo {author} {\bibfnamefont {E.}~\bibnamefont
  {Barausse}} \emph {et~al.},\ }\href {\doibase 10.1007/s10714-020-02691-1}
  {\bibfield  {journal} {\bibinfo  {journal} {General Relativity and
  Gravitation}\ }\textbf {\bibinfo {volume} {52}},\ \bibinfo {eid} {81}
  (\bibinfo {year} {2020})},\ \Eprint {http://arxiv.org/abs/2001.09793}
  {arXiv:2001.09793 [gr-qc]} \BibitemShut {NoStop}%
\bibitem [{\citenamefont {{Cai}}\ \emph {et~al.}(2017)\citenamefont {{Cai}},
  \citenamefont {{Tamanini}},\ and\ \citenamefont
  {{Yang}}}]{2017JCAP...05..031C}%
  \BibitemOpen
  \bibfield  {author} {\bibinfo {author} {\bibfnamefont {R.-G.}\ \bibnamefont
  {{Cai}}}, \bibinfo {author} {\bibfnamefont {N.}~\bibnamefont {{Tamanini}}}, \
  and\ \bibinfo {author} {\bibfnamefont {T.}~\bibnamefont {{Yang}}},\ }\href
  {\doibase 10.1088/1475-7516/2017/05/031} {\bibfield  {journal} {\bibinfo
  {journal} {Journal of Cosmology and Astroparticle Physics}\ }\textbf
  {\bibinfo {volume} {2017}},\ \bibinfo {eid} {031} (\bibinfo {year} {2017})},\
  \Eprint {http://arxiv.org/abs/1703.07323} {arXiv:1703.07323 [astro-ph.CO]}
  \BibitemShut {NoStop}%
\bibitem [{\citenamefont {Barack}\ \emph {et~al.}(2019)\citenamefont {Barack}
  \emph {et~al.}}]{2019CQGra..36n3001B}%
  \BibitemOpen
  \bibfield  {author} {\bibinfo {author} {\bibfnamefont {L.}~\bibnamefont
  {Barack}} \emph {et~al.},\ }\href {\doibase 10.1088/1361-6382/ab0587}
  {\bibfield  {journal} {\bibinfo  {journal} {Classical and Quantum Gravity}\
  }\textbf {\bibinfo {volume} {36}},\ \bibinfo {eid} {143001} (\bibinfo {year}
  {2019})},\ \Eprint {http://arxiv.org/abs/1806.05195} {arXiv:1806.05195
  [gr-qc]} \BibitemShut {NoStop}%
\bibitem [{\citenamefont {{Zhu}}\ \emph {et~al.}(2022)\citenamefont {{Zhu}}
  \emph {et~al.}}]{2022SCPMA..6559811Z}%
  \BibitemOpen
  \bibfield  {author} {\bibinfo {author} {\bibfnamefont {L.-G.}\ \bibnamefont
  {{Zhu}}} \emph {et~al.},\ }\href {\doibase 10.1007/s11433-021-1859-9}
  {\bibfield  {journal} {\bibinfo  {journal} {Science China Physics, Mechanics,
  and Astronomy}\ }\textbf {\bibinfo {volume} {65}},\ \bibinfo {eid} {259811}
  (\bibinfo {year} {2022})},\ \Eprint {http://arxiv.org/abs/2110.05224}
  {arXiv:2110.05224 [astro-ph.CO]} \BibitemShut {NoStop}%
\bibitem [{\citenamefont {{Kyutoku}}\ and\ \citenamefont
  {{Seto}}(2017)}]{2017PhRvD..95h3525K}%
  \BibitemOpen
  \bibfield  {author} {\bibinfo {author} {\bibfnamefont {K.}~\bibnamefont
  {{Kyutoku}}}\ and\ \bibinfo {author} {\bibfnamefont {N.}~\bibnamefont
  {{Seto}}},\ }\href {\doibase 10.1103/PhysRevD.95.083525} {\bibfield
  {journal} {\bibinfo  {journal} {\prd}\ }\textbf {\bibinfo {volume} {95}},\
  \bibinfo {eid} {083525} (\bibinfo {year} {2017})},\ \Eprint
  {http://arxiv.org/abs/1609.07142} {arXiv:1609.07142 [astro-ph.CO]}
  \BibitemShut {NoStop}%
\bibitem [{\citenamefont {{Trott}}\ and\ \citenamefont
  {{Huterer}}(2021)}]{2021arXiv211200241T}%
  \BibitemOpen
  \bibfield  {author} {\bibinfo {author} {\bibfnamefont {E.}~\bibnamefont
  {{Trott}}}\ and\ \bibinfo {author} {\bibfnamefont {D.}~\bibnamefont
  {{Huterer}}},\ }\href@noop {} {\bibfield  {journal} {\bibinfo  {journal}
  {arXiv e-prints}\ ,\ \bibinfo {eid} {arXiv:2112.00241}} (\bibinfo {year}
  {2021})},\ \Eprint {http://arxiv.org/abs/2112.00241} {arXiv:2112.00241
  [astro-ph.CO]} \BibitemShut {NoStop}%
\bibitem [{\citenamefont {{Belgacem}}\ \emph {et~al.}(2019)\citenamefont
  {{Belgacem}} \emph {et~al.}}]{2019JCAP...07..024B}%
  \BibitemOpen
  \bibfield  {author} {\bibinfo {author} {\bibfnamefont {E.}~\bibnamefont
  {{Belgacem}}} \emph {et~al.},\ }\href {\doibase
  10.1088/1475-7516/2019/07/024} {\bibfield  {journal} {\bibinfo  {journal}
  {Journal of Cosmology and Astroparticle Physics}\ }\textbf {\bibinfo {volume}
  {2019}},\ \bibinfo {eid} {024} (\bibinfo {year} {2019})},\ \Eprint
  {http://arxiv.org/abs/1906.01593} {arXiv:1906.01593 [astro-ph.CO]}
  \BibitemShut {NoStop}%
\bibitem [{\citenamefont {{Belgacem}}\ \emph {et~al.}(2018)\citenamefont
  {{Belgacem}}, \citenamefont {{Dirian}}, \citenamefont {{Foffa}},\ and\
  \citenamefont {{Maggiore}}}]{2018PhRvD..98b3510B}%
  \BibitemOpen
  \bibfield  {author} {\bibinfo {author} {\bibfnamefont {E.}~\bibnamefont
  {{Belgacem}}}, \bibinfo {author} {\bibfnamefont {Y.}~\bibnamefont
  {{Dirian}}}, \bibinfo {author} {\bibfnamefont {S.}~\bibnamefont {{Foffa}}}, \
  and\ \bibinfo {author} {\bibfnamefont {M.}~\bibnamefont {{Maggiore}}},\
  }\href {\doibase 10.1103/PhysRevD.98.023510} {\bibfield  {journal} {\bibinfo
  {journal} {\prd}\ }\textbf {\bibinfo {volume} {98}},\ \bibinfo {eid} {023510}
  (\bibinfo {year} {2018})},\ \Eprint {http://arxiv.org/abs/1805.08731}
  {arXiv:1805.08731 [gr-qc]} \BibitemShut {NoStop}%
\bibitem [{\citenamefont {{Speri}}\ \emph {et~al.}(2021)\citenamefont
  {{Speri}}, \citenamefont {{Tamanini}}, \citenamefont {{Caldwell}},
  \citenamefont {{Gair}},\ and\ \citenamefont {{Wang}}}]{2021PhRvD.103h3526S}%
  \BibitemOpen
  \bibfield  {author} {\bibinfo {author} {\bibfnamefont {L.}~\bibnamefont
  {{Speri}}}, \bibinfo {author} {\bibfnamefont {N.}~\bibnamefont {{Tamanini}}},
  \bibinfo {author} {\bibfnamefont {R.~R.}\ \bibnamefont {{Caldwell}}},
  \bibinfo {author} {\bibfnamefont {J.~R.}\ \bibnamefont {{Gair}}}, \ and\
  \bibinfo {author} {\bibfnamefont {B.}~\bibnamefont {{Wang}}},\ }\href
  {\doibase 10.1103/PhysRevD.103.083526} {\bibfield  {journal} {\bibinfo
  {journal} {\prd}\ }\textbf {\bibinfo {volume} {103}},\ \bibinfo {eid}
  {083526} (\bibinfo {year} {2021})},\ \Eprint
  {http://arxiv.org/abs/2010.09049} {arXiv:2010.09049 [astro-ph.CO]}
  \BibitemShut {NoStop}%
\bibitem [{\citenamefont {{Zel'dovich}}\ and\ \citenamefont
  {{Novikov}}(1967)}]{1967SvA....10..602Z}%
  \BibitemOpen
  \bibfield  {author} {\bibinfo {author} {\bibfnamefont {Y.~B.}\ \bibnamefont
  {{Zel'dovich}}}\ and\ \bibinfo {author} {\bibfnamefont {I.~D.}\ \bibnamefont
  {{Novikov}}},\ }\href@noop {} {\bibfield  {journal} {\bibinfo  {journal}
  {Soviet Astronomy}\ }\textbf {\bibinfo {volume} {10}},\ \bibinfo {pages}
  {602} (\bibinfo {year} {1967})}\BibitemShut {NoStop}%
\bibitem [{\citenamefont {{Carr}}\ and\ \citenamefont
  {{Hawking}}(1974)}]{Carr1974nx}%
  \BibitemOpen
  \bibfield  {author} {\bibinfo {author} {\bibfnamefont {B.~J.}\ \bibnamefont
  {{Carr}}}\ and\ \bibinfo {author} {\bibfnamefont {S.~W.}\ \bibnamefont
  {{Hawking}}},\ }\href@noop {} {\bibfield  {journal} {\bibinfo  {journal}
  {Mon. Not. Roy. Astron. Soc.}\ }\textbf {\bibinfo {volume} {168}},\ \bibinfo
  {pages} {399} (\bibinfo {year} {1974})}\BibitemShut {NoStop}%
\bibitem [{\citenamefont {{Carr}}(1975)}]{1975ApJ...201....1C}%
  \BibitemOpen
  \bibfield  {author} {\bibinfo {author} {\bibfnamefont {B.~J.}\ \bibnamefont
  {{Carr}}},\ }\href {\doibase 10.1086/153853} {\bibfield  {journal} {\bibinfo
  {journal} {ApJ}\ }\textbf {\bibinfo {volume} {201}},\ \bibinfo {pages} {1}
  (\bibinfo {year} {1975})}\BibitemShut {NoStop}%
\bibitem [{\citenamefont {Chapline}(1975)}]{Chapline1975ojl}%
  \BibitemOpen
  \bibfield  {author} {\bibinfo {author} {\bibfnamefont {G.~F.}\ \bibnamefont
  {Chapline}},\ }\href {\doibase 10.1038/253251a0} {\bibfield  {journal}
  {\bibinfo  {journal} {Nature}\ }\textbf {\bibinfo {volume} {253}},\ \bibinfo
  {pages} {251} (\bibinfo {year} {1975})}\BibitemShut {NoStop}%
\bibitem [{\citenamefont {Clesse}\ and\ \citenamefont
  {Garc\'\i{}a-Bellido}(2018)}]{Clesse2017bsw}%
  \BibitemOpen
  \bibfield  {author} {\bibinfo {author} {\bibfnamefont {S.}~\bibnamefont
  {Clesse}}\ and\ \bibinfo {author} {\bibfnamefont {J.}~\bibnamefont
  {Garc\'\i{}a-Bellido}},\ }\href {\doibase 10.1016/j.dark.2018.08.004}
  {\bibfield  {journal} {\bibinfo  {journal} {Phys. Dark Univ.}\ }\textbf
  {\bibinfo {volume} {22}},\ \bibinfo {pages} {137} (\bibinfo {year} {2018})},\
  \Eprint {http://arxiv.org/abs/1711.10458} {arXiv:1711.10458 [astro-ph.CO]}
  \BibitemShut {NoStop}%
\bibitem [{\citenamefont {Nakamura}\ \emph {et~al.}(1997)\citenamefont
  {Nakamura}, \citenamefont {Sasaki}, \citenamefont {Tanaka},\ and\
  \citenamefont {Thorne}}]{Nakamura1997sm}%
  \BibitemOpen
  \bibfield  {author} {\bibinfo {author} {\bibfnamefont {T.}~\bibnamefont
  {Nakamura}}, \bibinfo {author} {\bibfnamefont {M.}~\bibnamefont {Sasaki}},
  \bibinfo {author} {\bibfnamefont {T.}~\bibnamefont {Tanaka}}, \ and\ \bibinfo
  {author} {\bibfnamefont {K.~S.}\ \bibnamefont {Thorne}},\ }\href {\doibase
  10.1086/310886} {\bibfield  {journal} {\bibinfo  {journal} {Astrophys. J.}\
  }\textbf {\bibinfo {volume} {487}},\ \bibinfo {pages} {L139} (\bibinfo {year}
  {1997})},\ \Eprint {http://arxiv.org/abs/astro-ph/9708060}
  {arXiv:astro-ph/9708060 [astro-ph]} \BibitemShut {NoStop}%
%%CITATION = ASTRO-PH/9708060;%%
\bibitem [{\citenamefont {Ioka}\ \emph {et~al.}(1998)\citenamefont {Ioka},
  \citenamefont {Chiba}, \citenamefont {Tanaka},\ and\ \citenamefont
  {Nakamura}}]{Ioka1998nz}%
  \BibitemOpen
  \bibfield  {author} {\bibinfo {author} {\bibfnamefont {K.}~\bibnamefont
  {Ioka}}, \bibinfo {author} {\bibfnamefont {T.}~\bibnamefont {Chiba}},
  \bibinfo {author} {\bibfnamefont {T.}~\bibnamefont {Tanaka}}, \ and\ \bibinfo
  {author} {\bibfnamefont {T.}~\bibnamefont {Nakamura}},\ }\href {\doibase
  10.1103/PhysRevD.58.063003} {\bibfield  {journal} {\bibinfo  {journal} {Phys.
  Rev.}\ }\textbf {\bibinfo {volume} {D58}},\ \bibinfo {pages} {063003}
  (\bibinfo {year} {1998})},\ \Eprint {http://arxiv.org/abs/astro-ph/9807018}
  {arXiv:astro-ph/9807018 [astro-ph]} \BibitemShut {NoStop}%
%%CITATION = ASTRO-PH/9807018;%%
\bibitem [{\citenamefont {Eroshenko}(2018)}]{Eroshenko2016hmn}%
  \BibitemOpen
  \bibfield  {author} {\bibinfo {author} {\bibfnamefont {Y.~N.}\ \bibnamefont
  {Eroshenko}},\ }\href {\doibase 10.1088/1742-6596/1051/1/012010} {\bibfield
  {journal} {\bibinfo  {journal} {J. Phys. Conf. Ser.}\ }\textbf {\bibinfo
  {volume} {1051}},\ \bibinfo {pages} {012010} (\bibinfo {year} {2018})},\
  \Eprint {http://arxiv.org/abs/1604.04932} {arXiv:1604.04932 [astro-ph.CO]}
  \BibitemShut {NoStop}%
\bibitem [{\citenamefont {Raidal}\ \emph {et~al.}(2017)\citenamefont {Raidal},
  \citenamefont {Vaskonen},\ and\ \citenamefont {Veerm\"ae}}]{Raidal2017mfl}%
  \BibitemOpen
  \bibfield  {author} {\bibinfo {author} {\bibfnamefont {M.}~\bibnamefont
  {Raidal}}, \bibinfo {author} {\bibfnamefont {V.}~\bibnamefont {Vaskonen}}, \
  and\ \bibinfo {author} {\bibfnamefont {H.}~\bibnamefont {Veerm\"ae}},\ }\href
  {\doibase 10.1088/1475-7516/2017/09/037} {\bibfield  {journal} {\bibinfo
  {journal} {JCAP}\ }\textbf {\bibinfo {volume} {09}},\ \bibinfo {pages} {037}
  (\bibinfo {year} {2017})},\ \Eprint {http://arxiv.org/abs/1707.01480}
  {arXiv:1707.01480 [astro-ph.CO]} \BibitemShut {NoStop}%
\bibitem [{\citenamefont {Zagorac}\ \emph {et~al.}(2019)\citenamefont
  {Zagorac}, \citenamefont {Easther},\ and\ \citenamefont
  {Padmanabhan}}]{Zagorac2019ekv}%
  \BibitemOpen
  \bibfield  {author} {\bibinfo {author} {\bibfnamefont {J.~L.}\ \bibnamefont
  {Zagorac}}, \bibinfo {author} {\bibfnamefont {R.}~\bibnamefont {Easther}}, \
  and\ \bibinfo {author} {\bibfnamefont {N.}~\bibnamefont {Padmanabhan}},\
  }\href {\doibase 10.1088/1475-7516/2019/06/052} {\bibfield  {journal}
  {\bibinfo  {journal} {JCAP}\ }\textbf {\bibinfo {volume} {1906}},\ \bibinfo
  {pages} {052} (\bibinfo {year} {2019})},\ \Eprint
  {http://arxiv.org/abs/1903.05053} {arXiv:1903.05053 [astro-ph.CO]}
  \BibitemShut {NoStop}%
%%CITATION = ARXIV:1903.05053;%%
\bibitem [{\citenamefont {Hooper}\ \emph {et~al.}(2020)\citenamefont {Hooper},
  \citenamefont {Krnjaic}, \citenamefont {March-Russell}, \citenamefont
  {McDermott},\ and\ \citenamefont {Petrossian-Byrne}}]{Hooper2020evu}%
  \BibitemOpen
  \bibfield  {author} {\bibinfo {author} {\bibfnamefont {D.}~\bibnamefont
  {Hooper}}, \bibinfo {author} {\bibfnamefont {G.}~\bibnamefont {Krnjaic}},
  \bibinfo {author} {\bibfnamefont {J.}~\bibnamefont {March-Russell}}, \bibinfo
  {author} {\bibfnamefont {S.~D.}\ \bibnamefont {McDermott}}, \ and\ \bibinfo
  {author} {\bibfnamefont {R.}~\bibnamefont {Petrossian-Byrne}},\ }\href@noop
  {} {\  (\bibinfo {year} {2020})},\ \Eprint {http://arxiv.org/abs/2004.00618}
  {arXiv:2004.00618 [astro-ph.CO]} \BibitemShut {NoStop}%
\bibitem [{\citenamefont {Anantua}\ \emph {et~al.}(2009)\citenamefont
  {Anantua}, \citenamefont {Easther},\ and\ \citenamefont
  {Giblin}}]{Anantua2008am}%
  \BibitemOpen
  \bibfield  {author} {\bibinfo {author} {\bibfnamefont {R.}~\bibnamefont
  {Anantua}}, \bibinfo {author} {\bibfnamefont {R.}~\bibnamefont {Easther}}, \
  and\ \bibinfo {author} {\bibfnamefont {J.~T.}\ \bibnamefont {Giblin}},\
  }\href {\doibase 10.1103/PhysRevLett.103.111303} {\bibfield  {journal}
  {\bibinfo  {journal} {Phys. Rev. Lett.}\ }\textbf {\bibinfo {volume} {103}},\
  \bibinfo {pages} {111303} (\bibinfo {year} {2009})},\ \Eprint
  {http://arxiv.org/abs/0812.0825} {arXiv:0812.0825 [astro-ph]} \BibitemShut
  {NoStop}%
\bibitem [{\citenamefont {Dong}\ \emph {et~al.}(2016)\citenamefont {Dong},
  \citenamefont {Kinney},\ and\ \citenamefont {Stojkovic}}]{Dong2015yjs}%
  \BibitemOpen
  \bibfield  {author} {\bibinfo {author} {\bibfnamefont {R.}~\bibnamefont
  {Dong}}, \bibinfo {author} {\bibfnamefont {W.~H.}\ \bibnamefont {Kinney}}, \
  and\ \bibinfo {author} {\bibfnamefont {D.}~\bibnamefont {Stojkovic}},\ }\href
  {\doibase 10.1088/1475-7516/2016/10/034} {\bibfield  {journal} {\bibinfo
  {journal} {JCAP}\ }\textbf {\bibinfo {volume} {10}},\ \bibinfo {pages} {034}
  (\bibinfo {year} {2016})},\ \Eprint {http://arxiv.org/abs/1511.05642}
  {arXiv:1511.05642 [astro-ph.CO]} \BibitemShut {NoStop}%
\bibitem [{\citenamefont {Dom\`enech}(2021)}]{Domenech2021ztg}%
  \BibitemOpen
  \bibfield  {author} {\bibinfo {author} {\bibfnamefont {G.}~\bibnamefont
  {Dom\`enech}},\ }\href {\doibase 10.3390/universe7110398} {\bibfield
  {journal} {\bibinfo  {journal} {Universe}\ }\textbf {\bibinfo {volume} {7}},\
  \bibinfo {pages} {398} (\bibinfo {year} {2021})},\ \Eprint
  {http://arxiv.org/abs/2109.01398} {arXiv:2109.01398 [gr-qc]} \BibitemShut
  {NoStop}%
\bibitem [{\citenamefont {Papanikolaou}\ \emph {et~al.}(2021)\citenamefont
  {Papanikolaou}, \citenamefont {Vennin},\ and\ \citenamefont
  {Langlois}}]{Papanikolaou2020qtd}%
  \BibitemOpen
  \bibfield  {author} {\bibinfo {author} {\bibfnamefont {T.}~\bibnamefont
  {Papanikolaou}}, \bibinfo {author} {\bibfnamefont {V.}~\bibnamefont
  {Vennin}}, \ and\ \bibinfo {author} {\bibfnamefont {D.}~\bibnamefont
  {Langlois}},\ }\href {\doibase 10.1088/1475-7516/2021/03/053} {\bibfield
  {journal} {\bibinfo  {journal} {JCAP}\ }\textbf {\bibinfo {volume} {03}},\
  \bibinfo {pages} {053} (\bibinfo {year} {2021})},\ \Eprint
  {http://arxiv.org/abs/2010.11573} {arXiv:2010.11573 [astro-ph.CO]}
  \BibitemShut {NoStop}%
\bibitem [{\citenamefont {Papanikolaou}(2022)}]{Papanikolaou2022chm}%
  \BibitemOpen
  \bibfield  {author} {\bibinfo {author} {\bibfnamefont {T.}~\bibnamefont
  {Papanikolaou}},\ }\href {\doibase 10.1088/1475-7516/2022/10/089} {\bibfield
  {journal} {\bibinfo  {journal} {JCAP}\ }\textbf {\bibinfo {volume} {10}},\
  \bibinfo {pages} {089} (\bibinfo {year} {2022})},\ \Eprint
  {http://arxiv.org/abs/2207.11041} {arXiv:2207.11041 [astro-ph.CO]}
  \BibitemShut {NoStop}%
\bibitem [{\citenamefont {Bartolo}\ \emph {et~al.}(2019)\citenamefont
  {Bartolo}, \citenamefont {De~Luca}, \citenamefont {Franciolini},
  \citenamefont {Lewis}, \citenamefont {Peloso},\ and\ \citenamefont
  {Riotto}}]{Bartolo2018evs}%
  \BibitemOpen
  \bibfield  {author} {\bibinfo {author} {\bibfnamefont {N.}~\bibnamefont
  {Bartolo}}, \bibinfo {author} {\bibfnamefont {V.}~\bibnamefont {De~Luca}},
  \bibinfo {author} {\bibfnamefont {G.}~\bibnamefont {Franciolini}}, \bibinfo
  {author} {\bibfnamefont {A.}~\bibnamefont {Lewis}}, \bibinfo {author}
  {\bibfnamefont {M.}~\bibnamefont {Peloso}}, \ and\ \bibinfo {author}
  {\bibfnamefont {A.}~\bibnamefont {Riotto}},\ }\href {\doibase
  10.1103/PhysRevLett.122.211301} {\bibfield  {journal} {\bibinfo  {journal}
  {Phys. Rev. Lett.}\ }\textbf {\bibinfo {volume} {122}},\ \bibinfo {pages}
  {211301} (\bibinfo {year} {2019})},\ \Eprint
  {http://arxiv.org/abs/1810.12218} {arXiv:1810.12218 [astro-ph.CO]}
  \BibitemShut {NoStop}%
\bibitem [{\citenamefont {Cai}\ \emph {et~al.}(2019{\natexlab{a}})\citenamefont
  {Cai}, \citenamefont {Pi},\ and\ \citenamefont {Sasaki}}]{Cai}%
  \BibitemOpen
  \bibfield  {author} {\bibinfo {author} {\bibfnamefont {R.-G.}\ \bibnamefont
  {Cai}}, \bibinfo {author} {\bibfnamefont {S.}~\bibnamefont {Pi}}, \ and\
  \bibinfo {author} {\bibfnamefont {M.}~\bibnamefont {Sasaki}},\ }\href
  {\doibase 10.1103/PhysRevLett.122.201101} {\bibfield  {journal} {\bibinfo
  {journal} {Phys. Rev. Lett.}\ }\textbf {\bibinfo {volume} {122}},\ \bibinfo
  {pages} {201101} (\bibinfo {year} {2019}{\natexlab{a}})}\BibitemShut
  {NoStop}%
\bibitem [{\citenamefont {Karnesis}\ \emph {et~al.}(2020)\citenamefont
  {Karnesis}, \citenamefont {Lilley},\ and\ \citenamefont
  {Petiteau}}]{Karnesis2019mph}%
  \BibitemOpen
  \bibfield  {author} {\bibinfo {author} {\bibfnamefont {N.}~\bibnamefont
  {Karnesis}}, \bibinfo {author} {\bibfnamefont {M.}~\bibnamefont {Lilley}}, \
  and\ \bibinfo {author} {\bibfnamefont {A.}~\bibnamefont {Petiteau}},\ }\href
  {\doibase 10.1088/1361-6382/abb637} {\bibfield  {journal} {\bibinfo
  {journal} {Class. Quant. Grav.}\ }\textbf {\bibinfo {volume} {37}},\ \bibinfo
  {pages} {215017} (\bibinfo {year} {2020})},\ \Eprint
  {http://arxiv.org/abs/1906.09027} {arXiv:1906.09027 [astro-ph.IM]}
  \BibitemShut {NoStop}%
\bibitem [{\citenamefont {Flauger}\ \emph {et~al.}(2020)\citenamefont
  {Flauger}, \citenamefont {Karnesis}, \citenamefont {Nardini}, \citenamefont
  {Pieroni}, \citenamefont {Ricciardone},\ and\ \citenamefont
  {Torrado}}]{Flauger2020qyi}%
  \BibitemOpen
  \bibfield  {author} {\bibinfo {author} {\bibfnamefont {R.}~\bibnamefont
  {Flauger}}, \bibinfo {author} {\bibfnamefont {N.}~\bibnamefont {Karnesis}},
  \bibinfo {author} {\bibfnamefont {G.}~\bibnamefont {Nardini}}, \bibinfo
  {author} {\bibfnamefont {M.}~\bibnamefont {Pieroni}}, \bibinfo {author}
  {\bibfnamefont {A.}~\bibnamefont {Ricciardone}}, \ and\ \bibinfo {author}
  {\bibfnamefont {J.}~\bibnamefont {Torrado}},\ }\href@noop {} {\  (\bibinfo
  {year} {2020})},\ \Eprint {http://arxiv.org/abs/2009.11845} {arXiv:2009.11845
  [astro-ph.CO]} \BibitemShut {NoStop}%
\bibitem [{\citenamefont {Baghi}\ \emph {et~al.}(2023)\citenamefont {Baghi},
  \citenamefont {Karnesis}, \citenamefont {Bayle}, \citenamefont
  {Besan\c{c}on},\ and\ \citenamefont {Inchausp\'e}}]{Baghi2023qnq}%
  \BibitemOpen
  \bibfield  {author} {\bibinfo {author} {\bibfnamefont {Q.}~\bibnamefont
  {Baghi}}, \bibinfo {author} {\bibfnamefont {N.}~\bibnamefont {Karnesis}},
  \bibinfo {author} {\bibfnamefont {J.-B.}\ \bibnamefont {Bayle}}, \bibinfo
  {author} {\bibfnamefont {M.}~\bibnamefont {Besan\c{c}on}}, \ and\ \bibinfo
  {author} {\bibfnamefont {H.}~\bibnamefont {Inchausp\'e}},\ }\href {\doibase
  10.1088/1475-7516/2023/04/066} {\bibfield  {journal} {\bibinfo  {journal}
  {JCAP}\ }\textbf {\bibinfo {volume} {04}},\ \bibinfo {pages} {066} (\bibinfo
  {year} {2023})},\ \Eprint {http://arxiv.org/abs/2302.12573} {arXiv:2302.12573
  [gr-qc]} \BibitemShut {NoStop}%
\bibitem [{\citenamefont {Cai}\ \emph {et~al.}(2019{\natexlab{b}})\citenamefont
  {Cai}, \citenamefont {Pi},\ and\ \citenamefont {Sasaki}}]{Cai2018dig}%
  \BibitemOpen
  \bibfield  {author} {\bibinfo {author} {\bibfnamefont {R.-g.}\ \bibnamefont
  {Cai}}, \bibinfo {author} {\bibfnamefont {S.}~\bibnamefont {Pi}}, \ and\
  \bibinfo {author} {\bibfnamefont {M.}~\bibnamefont {Sasaki}},\ }\href
  {\doibase 10.1103/PhysRevLett.122.201101} {\bibfield  {journal} {\bibinfo
  {journal} {Phys. Rev. Lett.}\ }\textbf {\bibinfo {volume} {122}},\ \bibinfo
  {pages} {201101} (\bibinfo {year} {2019}{\natexlab{b}})},\ \Eprint
  {http://arxiv.org/abs/1810.11000} {arXiv:1810.11000 [astro-ph.CO]}
  \BibitemShut {NoStop}%
\bibitem [{\citenamefont {Papanikolaou}\ \emph
  {et~al.}(2023{\natexlab{a}})\citenamefont {Papanikolaou}, \citenamefont
  {Lymperis}, \citenamefont {Lola},\ and\ \citenamefont
  {Saridakis}}]{Papanikolaou2022did}%
  \BibitemOpen
  \bibfield  {author} {\bibinfo {author} {\bibfnamefont {T.}~\bibnamefont
  {Papanikolaou}}, \bibinfo {author} {\bibfnamefont {A.}~\bibnamefont
  {Lymperis}}, \bibinfo {author} {\bibfnamefont {S.}~\bibnamefont {Lola}}, \
  and\ \bibinfo {author} {\bibfnamefont {E.~N.}\ \bibnamefont {Saridakis}},\
  }\href {\doibase 10.1088/1475-7516/2023/03/003} {\bibfield  {journal}
  {\bibinfo  {journal} {JCAP}\ }\textbf {\bibinfo {volume} {03}},\ \bibinfo
  {pages} {003} (\bibinfo {year} {2023}{\natexlab{a}})},\ \Eprint
  {http://arxiv.org/abs/2211.14900} {arXiv:2211.14900 [astro-ph.CO]}
  \BibitemShut {NoStop}%
\bibitem [{\citenamefont {Basilakos}\ \emph {et~al.}(2023)\citenamefont
  {Basilakos}, \citenamefont {Nanopoulos}, \citenamefont {Papanikolaou},
  \citenamefont {Saridakis},\ and\ \citenamefont
  {Tzerefos}}]{Basilakos2023xof}%
  \BibitemOpen
  \bibfield  {author} {\bibinfo {author} {\bibfnamefont {S.}~\bibnamefont
  {Basilakos}}, \bibinfo {author} {\bibfnamefont {D.~V.}\ \bibnamefont
  {Nanopoulos}}, \bibinfo {author} {\bibfnamefont {T.}~\bibnamefont
  {Papanikolaou}}, \bibinfo {author} {\bibfnamefont {E.~N.}\ \bibnamefont
  {Saridakis}}, \ and\ \bibinfo {author} {\bibfnamefont {C.}~\bibnamefont
  {Tzerefos}},\ }\href@noop {} {\  (\bibinfo {year} {2023})},\ \Eprint
  {http://arxiv.org/abs/2307.08601} {arXiv:2307.08601 [hep-th]} \BibitemShut
  {NoStop}%
\bibitem [{\citenamefont {Papanikolaou}\ \emph {et~al.}(2022)\citenamefont
  {Papanikolaou}, \citenamefont {Tzerefos}, \citenamefont {Basilakos},\ and\
  \citenamefont {Saridakis}}]{Papanikolaou2021uhe}%
  \BibitemOpen
  \bibfield  {author} {\bibinfo {author} {\bibfnamefont {T.}~\bibnamefont
  {Papanikolaou}}, \bibinfo {author} {\bibfnamefont {C.}~\bibnamefont
  {Tzerefos}}, \bibinfo {author} {\bibfnamefont {S.}~\bibnamefont {Basilakos}},
  \ and\ \bibinfo {author} {\bibfnamefont {E.~N.}\ \bibnamefont {Saridakis}},\
  }\href {\doibase 10.1088/1475-7516/2022/10/013} {\bibfield  {journal}
  {\bibinfo  {journal} {JCAP}\ }\textbf {\bibinfo {volume} {10}},\ \bibinfo
  {pages} {013} (\bibinfo {year} {2022})},\ \Eprint
  {http://arxiv.org/abs/2112.15059} {arXiv:2112.15059 [astro-ph.CO]}
  \BibitemShut {NoStop}%
\bibitem [{\citenamefont {Papanikolaou}\ \emph
  {et~al.}(2023{\natexlab{b}})\citenamefont {Papanikolaou}, \citenamefont
  {Tzerefos}, \citenamefont {Basilakos},\ and\ \citenamefont
  {Saridakis}}]{Papanikolaou2022hkg}%
  \BibitemOpen
  \bibfield  {author} {\bibinfo {author} {\bibfnamefont {T.}~\bibnamefont
  {Papanikolaou}}, \bibinfo {author} {\bibfnamefont {C.}~\bibnamefont
  {Tzerefos}}, \bibinfo {author} {\bibfnamefont {S.}~\bibnamefont {Basilakos}},
  \ and\ \bibinfo {author} {\bibfnamefont {E.~N.}\ \bibnamefont {Saridakis}},\
  }\href {\doibase 10.1140/epjc/s10052-022-11157-4} {\bibfield  {journal}
  {\bibinfo  {journal} {Eur. Phys. J. C}\ }\textbf {\bibinfo {volume} {83}},\
  \bibinfo {pages} {31} (\bibinfo {year} {2023}{\natexlab{b}})},\ \Eprint
  {http://arxiv.org/abs/2205.06094} {arXiv:2205.06094 [gr-qc]} \BibitemShut
  {NoStop}%
\bibitem [{\citenamefont {Auclair}\ \emph {et~al.}(2022)\citenamefont {Auclair}
  \emph {et~al.}}]{LISACosmologyWorkingGroup2022jok}%
  \BibitemOpen
  \bibfield  {author} {\bibinfo {author} {\bibfnamefont {P.}~\bibnamefont
  {Auclair}} \emph {et~al.} (\bibinfo {collaboration} {LISA Cosmology Working
  Group}),\ }\href@noop {} {\  (\bibinfo {year} {2022})},\ \Eprint
  {http://arxiv.org/abs/2204.05434} {arXiv:2204.05434 [astro-ph.CO]}
  \BibitemShut {NoStop}%
\bibitem [{\citenamefont {Abbott}\ \emph
  {et~al.}(2020{\natexlab{b}})\citenamefont {Abbott} \emph
  {et~al.}}]{LIGOScientific2020zkf}%
  \BibitemOpen
  \bibfield  {author} {\bibinfo {author} {\bibfnamefont {R.}~\bibnamefont
  {Abbott}} \emph {et~al.} (\bibinfo {collaboration} {LIGO Scientific,
  Virgo}),\ }\href {\doibase 10.3847/2041-8213/ab960f} {\bibfield  {journal}
  {\bibinfo  {journal} {Astrophys. J. Lett.}\ }\textbf {\bibinfo {volume}
  {896}},\ \bibinfo {pages} {L44} (\bibinfo {year} {2020}{\natexlab{b}})},\
  \Eprint {http://arxiv.org/abs/2006.12611} {arXiv:2006.12611 [astro-ph.HE]}
  \BibitemShut {NoStop}%
\bibitem [{\citenamefont {Anastopoulos}\ and\ \citenamefont
  {Hu}(2013)}]{AHGravDec}%
  \BibitemOpen
  \bibfield  {author} {\bibinfo {author} {\bibfnamefont {C.}~\bibnamefont
  {Anastopoulos}}\ and\ \bibinfo {author} {\bibfnamefont {B.~L.}\ \bibnamefont
  {Hu}},\ }\href@noop {} {\bibfield  {journal} {\bibinfo  {journal} {Classical
  and Quantum Gravity}\ }\textbf {\bibinfo {volume} {30}},\ \bibinfo {pages}
  {165007} (\bibinfo {year} {2013})}\BibitemShut {NoStop}%
\bibitem [{\citenamefont {Lagouvardos}\ and\ \citenamefont
  {Anastopoulos}(2021)}]{LagAn21}%
  \BibitemOpen
  \bibfield  {author} {\bibinfo {author} {\bibfnamefont {M.}~\bibnamefont
  {Lagouvardos}}\ and\ \bibinfo {author} {\bibfnamefont {C.}~\bibnamefont
  {Anastopoulos}},\ }\href {\doibase 10.1088/1361-6382/abf2f3} {\bibfield
  {journal} {\bibinfo  {journal} {Classical and Quantum Gravity}\ }\textbf
  {\bibinfo {volume} {38}},\ \bibinfo {pages} {115012} (\bibinfo {year}
  {2021})}\BibitemShut {NoStop}%
\bibitem [{\citenamefont {Parikh}\ \emph {et~al.}(2020)\citenamefont {Parikh},
  \citenamefont {Wilczek},\ and\ \citenamefont {Zahariade}}]{PWZ20}%
  \BibitemOpen
  \bibfield  {author} {\bibinfo {author} {\bibfnamefont {M.}~\bibnamefont
  {Parikh}}, \bibinfo {author} {\bibfnamefont {F.}~\bibnamefont {Wilczek}}, \
  and\ \bibinfo {author} {\bibfnamefont {G.}~\bibnamefont {Zahariade}},\ }\href
  {\doibase https://doi.org/10.1142/S0218271820420018} {\bibfield  {journal}
  {\bibinfo  {journal} {International Journal of Modern Physics D}\ }\textbf
  {\bibinfo {volume} {29}},\ \bibinfo {pages} {2042001} (\bibinfo {year}
  {2020})}\BibitemShut {NoStop}%
\bibitem [{\citenamefont {Mohageg}\ \emph {et~al.}(2022)\citenamefont {Mohageg}
  \emph {et~al.}}]{DSQL}%
  \BibitemOpen
  \bibfield  {author} {\bibinfo {author} {\bibfnamefont {M.}~\bibnamefont
  {Mohageg}} \emph {et~al.},\ }\href {\doibase
  10.1140/epjqt/s40507-022-00143-0} {\bibfield  {journal} {\bibinfo  {journal}
  {EPJ Quant. Technol.}\ }\textbf {\bibinfo {volume} {9}},\ \bibinfo {pages}
  {25} (\bibinfo {year} {2022})},\ \Eprint {http://arxiv.org/abs/2111.15591}
  {arXiv:2111.15591 [quant-ph]} \BibitemShut {NoStop}%
\bibitem [{\citenamefont {Saridakis}\ \emph {et~al.}(2021)\citenamefont
  {Saridakis} \emph {et~al.}}]{CANTATA2021ktz}%
  \BibitemOpen
  \bibfield  {author} {\bibinfo {author} {\bibfnamefont {E.~N.}\ \bibnamefont
  {Saridakis}} \emph {et~al.} (\bibinfo {collaboration} {CANTATA}),\
  }\href@noop {} {\  (\bibinfo {year} {2021})},\ \Eprint
  {http://arxiv.org/abs/2105.12582} {arXiv:2105.12582 [gr-qc]} \BibitemShut
  {NoStop}%
\bibitem [{\citenamefont {{Addazi}}\ \emph {et~al.}(2022)\citenamefont
  {{Addazi}} \emph {et~al.}}]{2022PrPNP.12503948A}%
  \BibitemOpen
  \bibfield  {author} {\bibinfo {author} {\bibfnamefont {A.}~\bibnamefont
  {{Addazi}}} \emph {et~al.},\ }\href {\doibase 10.1016/j.ppnp.2022.103948}
  {\bibfield  {journal} {\bibinfo  {journal} {Progress in Particle and Nuclear
  Physics}\ }\textbf {\bibinfo {volume} {125}},\ \bibinfo {eid} {103948}
  (\bibinfo {year} {2022})},\ \Eprint {http://arxiv.org/abs/2111.05659}
  {arXiv:2111.05659 [hep-ph]} \BibitemShut {NoStop}%
\bibitem [{\citenamefont {{Das}}\ and\ \citenamefont
  {{Vagenas}}(2008)}]{2008PhRvL.101v1301D}%
  \BibitemOpen
  \bibfield  {author} {\bibinfo {author} {\bibfnamefont {S.}~\bibnamefont
  {{Das}}}\ and\ \bibinfo {author} {\bibfnamefont {E.~C.}\ \bibnamefont
  {{Vagenas}}},\ }\href {\doibase 10.1103/PhysRevLett.101.221301} {\bibfield
  {journal} {\bibinfo  {journal} {\prl}\ }\textbf {\bibinfo {volume} {101}},\
  \bibinfo {eid} {221301} (\bibinfo {year} {2008})},\ \Eprint
  {http://arxiv.org/abs/0810.5333} {arXiv:0810.5333 [hep-th]} \BibitemShut
  {NoStop}%
\bibitem [{\citenamefont {{Das}}\ and\ \citenamefont
  {{Vagenas}}(2010)}]{2010PhRvL.104k9002D}%
  \BibitemOpen
  \bibfield  {author} {\bibinfo {author} {\bibfnamefont {S.}~\bibnamefont
  {{Das}}}\ and\ \bibinfo {author} {\bibfnamefont {E.~C.}\ \bibnamefont
  {{Vagenas}}},\ }\href {\doibase 10.1103/PhysRevLett.104.119002} {\bibfield
  {journal} {\bibinfo  {journal} {\prl}\ }\textbf {\bibinfo {volume} {104}},\
  \bibinfo {eid} {119002} (\bibinfo {year} {2010})},\ \Eprint
  {http://arxiv.org/abs/1003.3208} {arXiv:1003.3208 [hep-th]} \BibitemShut
  {NoStop}%
\bibitem [{\citenamefont {{Das}}\ and\ \citenamefont
  {{Vagenas}}(2009)}]{2009CaJPh..87..233D}%
  \BibitemOpen
  \bibfield  {author} {\bibinfo {author} {\bibfnamefont {S.}~\bibnamefont
  {{Das}}}\ and\ \bibinfo {author} {\bibfnamefont {E.~C.}\ \bibnamefont
  {{Vagenas}}},\ }\href {\doibase 10.1139/P08-105} {\bibfield  {journal}
  {\bibinfo  {journal} {Canadian Journal of Physics}\ }\textbf {\bibinfo
  {volume} {87}},\ \bibinfo {pages} {233} (\bibinfo {year} {2009})},\ \Eprint
  {http://arxiv.org/abs/0901.1768} {arXiv:0901.1768 [hep-th]} \BibitemShut
  {NoStop}%
\bibitem [{\citenamefont {{Ali}}\ \emph {et~al.}(2011)\citenamefont {{Ali}},
  \citenamefont {{Das}},\ and\ \citenamefont
  {{Vagenas}}}]{2011PhRvD..84d4013A}%
  \BibitemOpen
  \bibfield  {author} {\bibinfo {author} {\bibfnamefont {A.~F.}\ \bibnamefont
  {{Ali}}}, \bibinfo {author} {\bibfnamefont {S.}~\bibnamefont {{Das}}}, \ and\
  \bibinfo {author} {\bibfnamefont {E.~C.}\ \bibnamefont {{Vagenas}}},\ }\href
  {\doibase 10.1103/PhysRevD.84.044013} {\bibfield  {journal} {\bibinfo
  {journal} {\prd}\ }\textbf {\bibinfo {volume} {84}},\ \bibinfo {eid} {044013}
  (\bibinfo {year} {2011})},\ \Eprint {http://arxiv.org/abs/1107.3164}
  {arXiv:1107.3164 [hep-th]} \BibitemShut {NoStop}%
\bibitem [{\citenamefont {{Ali}}\ \emph {et~al.}(2009)\citenamefont {{Ali}},
  \citenamefont {{Das}},\ and\ \citenamefont
  {{Vagenas}}}]{2009PhLB..678..497A}%
  \BibitemOpen
  \bibfield  {author} {\bibinfo {author} {\bibfnamefont {A.~F.}\ \bibnamefont
  {{Ali}}}, \bibinfo {author} {\bibfnamefont {S.}~\bibnamefont {{Das}}}, \ and\
  \bibinfo {author} {\bibfnamefont {E.~C.}\ \bibnamefont {{Vagenas}}},\ }\href
  {\doibase 10.1016/j.physletb.2009.06.061} {\bibfield  {journal} {\bibinfo
  {journal} {Physics Letters B}\ }\textbf {\bibinfo {volume} {678}},\ \bibinfo
  {pages} {497} (\bibinfo {year} {2009})},\ \Eprint
  {http://arxiv.org/abs/0906.5396} {arXiv:0906.5396 [hep-th]} \BibitemShut
  {NoStop}%
\bibitem [{\citenamefont {{Das}}\ \emph
  {et~al.}(2010{\natexlab{a}})\citenamefont {{Das}}, \citenamefont
  {{Vagenas}},\ and\ \citenamefont {{Ali}}}]{2010PhLB..690..407D}%
  \BibitemOpen
  \bibfield  {author} {\bibinfo {author} {\bibfnamefont {S.}~\bibnamefont
  {{Das}}}, \bibinfo {author} {\bibfnamefont {E.~C.}\ \bibnamefont
  {{Vagenas}}}, \ and\ \bibinfo {author} {\bibfnamefont {A.~F.}\ \bibnamefont
  {{Ali}}},\ }\href {\doibase 10.1016/j.physletb.2010.05.052} {\bibfield
  {journal} {\bibinfo  {journal} {Physics Letters B}\ }\textbf {\bibinfo
  {volume} {690}},\ \bibinfo {pages} {407} (\bibinfo {year}
  {2010}{\natexlab{a}})},\ \Eprint {http://arxiv.org/abs/1005.3368}
  {arXiv:1005.3368 [hep-th]} \BibitemShut {NoStop}%
\bibitem [{\citenamefont {{Das}}\ \emph
  {et~al.}(2010{\natexlab{b}})\citenamefont {{Das}}, \citenamefont
  {{Vagenas}},\ and\ \citenamefont {{Farag Ali}}}]{2010PhLB..692..342D}%
  \BibitemOpen
  \bibfield  {author} {\bibinfo {author} {\bibfnamefont {S.}~\bibnamefont
  {{Das}}}, \bibinfo {author} {\bibfnamefont {E.~C.}\ \bibnamefont
  {{Vagenas}}}, \ and\ \bibinfo {author} {\bibfnamefont {A.}~\bibnamefont
  {{Farag Ali}}},\ }\href {\doibase 10.1016/j.physletb.2010.07.025} {\bibfield
  {journal} {\bibinfo  {journal} {Physics Letters B}\ }\textbf {\bibinfo
  {volume} {692}},\ \bibinfo {pages} {342} (\bibinfo {year}
  {2010}{\natexlab{b}})}\BibitemShut {NoStop}%
\bibitem [{\citenamefont {{Scardigli}}\ \emph {et~al.}(2017)\citenamefont
  {{Scardigli}}, \citenamefont {{Lambiase}},\ and\ \citenamefont
  {{Vagenas}}}]{2017PhLB..767..242S}%
  \BibitemOpen
  \bibfield  {author} {\bibinfo {author} {\bibfnamefont {F.}~\bibnamefont
  {{Scardigli}}}, \bibinfo {author} {\bibfnamefont {G.}~\bibnamefont
  {{Lambiase}}}, \ and\ \bibinfo {author} {\bibfnamefont {E.~C.}\ \bibnamefont
  {{Vagenas}}},\ }\href {\doibase 10.1016/j.physletb.2017.01.054} {\bibfield
  {journal} {\bibinfo  {journal} {Physics Letters B}\ }\textbf {\bibinfo
  {volume} {767}},\ \bibinfo {pages} {242} (\bibinfo {year} {2017})},\ \Eprint
  {http://arxiv.org/abs/1611.01469} {arXiv:1611.01469 [hep-th]} \BibitemShut
  {NoStop}%
\bibitem [{\citenamefont {{Vagenas}}\ \emph {et~al.}(2017)\citenamefont
  {{Vagenas}}, \citenamefont {{Alsaleh}},\ and\ \citenamefont {{Farag
  Ali}}}]{2017EL....12040001V}%
  \BibitemOpen
  \bibfield  {author} {\bibinfo {author} {\bibfnamefont {E.~C.}\ \bibnamefont
  {{Vagenas}}}, \bibinfo {author} {\bibfnamefont {S.~M.}\ \bibnamefont
  {{Alsaleh}}}, \ and\ \bibinfo {author} {\bibfnamefont {A.}~\bibnamefont
  {{Farag Ali}}},\ }\href {\doibase 10.1209/0295-5075/120/40001} {\bibfield
  {journal} {\bibinfo  {journal} {EPL (Europhysics Letters)}\ }\textbf
  {\bibinfo {volume} {120}},\ \bibinfo {pages} {40001} (\bibinfo {year}
  {2017})}\BibitemShut {NoStop}%
\bibitem [{\citenamefont {{Vagenas}}\ \emph {et~al.}(2019)\citenamefont
  {{Vagenas}}, \citenamefont {{Ali}},\ and\ \citenamefont
  {{Alshal}}}]{2019PhRvD..99h4013V}%
  \BibitemOpen
  \bibfield  {author} {\bibinfo {author} {\bibfnamefont {E.~C.}\ \bibnamefont
  {{Vagenas}}}, \bibinfo {author} {\bibfnamefont {A.~F.}\ \bibnamefont
  {{Ali}}}, \ and\ \bibinfo {author} {\bibfnamefont {H.}~\bibnamefont
  {{Alshal}}},\ }\href {\doibase 10.1103/PhysRevD.99.084013} {\bibfield
  {journal} {\bibinfo  {journal} {\prd}\ }\textbf {\bibinfo {volume} {99}},\
  \bibinfo {eid} {084013} (\bibinfo {year} {2019})},\ \Eprint
  {http://arxiv.org/abs/1903.09634} {arXiv:1903.09634 [hep-th]} \BibitemShut
  {NoStop}%
\bibitem [{\citenamefont {{Das}}\ \emph
  {et~al.}(2021{\natexlab{a}})\citenamefont {{Das}}, \citenamefont {{Das}},
  \citenamefont {{Mansour}},\ and\ \citenamefont
  {{Vagenas}}}]{2021PhLB..81936429D}%
  \BibitemOpen
  \bibfield  {author} {\bibinfo {author} {\bibfnamefont {A.}~\bibnamefont
  {{Das}}}, \bibinfo {author} {\bibfnamefont {S.}~\bibnamefont {{Das}}},
  \bibinfo {author} {\bibfnamefont {N.~R.}\ \bibnamefont {{Mansour}}}, \ and\
  \bibinfo {author} {\bibfnamefont {E.~C.}\ \bibnamefont {{Vagenas}}},\ }\href
  {\doibase 10.1016/j.physletb.2021.136429} {\bibfield  {journal} {\bibinfo
  {journal} {Physics Letters B}\ }\textbf {\bibinfo {volume} {819}},\ \bibinfo
  {eid} {136429} (\bibinfo {year} {2021}{\natexlab{a}})},\ \Eprint
  {http://arxiv.org/abs/2101.03746} {arXiv:2101.03746 [gr-qc]} \BibitemShut
  {NoStop}%
\bibitem [{\citenamefont {{Basilakos}}\ \emph {et~al.}(2010)\citenamefont
  {{Basilakos}}, \citenamefont {{Das}},\ and\ \citenamefont
  {{Vagenas}}}]{2010JCAP...09..027B}%
  \BibitemOpen
  \bibfield  {author} {\bibinfo {author} {\bibfnamefont {S.}~\bibnamefont
  {{Basilakos}}}, \bibinfo {author} {\bibfnamefont {S.}~\bibnamefont {{Das}}},
  \ and\ \bibinfo {author} {\bibfnamefont {E.~C.}\ \bibnamefont {{Vagenas}}},\
  }\href {\doibase 10.1088/1475-7516/2010/09/027} {\bibfield  {journal}
  {\bibinfo  {journal} {JCAP}\ }\textbf {\bibinfo {volume} {2010}},\ \bibinfo
  {eid} {027} (\bibinfo {year} {2010})},\ \Eprint
  {http://arxiv.org/abs/1009.0365} {arXiv:1009.0365 [hep-th]} \BibitemShut
  {NoStop}%
\bibitem [{\citenamefont {{Aghababaei}}\ \emph {et~al.}(2021)\citenamefont
  {{Aghababaei}}, \citenamefont {{Moradpour}},\ and\ \citenamefont
  {{Vagenas}}}]{2021EPJP..136..997A}%
  \BibitemOpen
  \bibfield  {author} {\bibinfo {author} {\bibfnamefont {S.}~\bibnamefont
  {{Aghababaei}}}, \bibinfo {author} {\bibfnamefont {H.}~\bibnamefont
  {{Moradpour}}}, \ and\ \bibinfo {author} {\bibfnamefont {E.~C.}\ \bibnamefont
  {{Vagenas}}},\ }\href {\doibase 10.1140/epjp/s13360-021-02007-5} {\bibfield
  {journal} {\bibinfo  {journal} {European Physical Journal Plus}\ }\textbf
  {\bibinfo {volume} {136}},\ \bibinfo {eid} {997} (\bibinfo {year}
  {2021})}\BibitemShut {NoStop}%
\bibitem [{\citenamefont {{Das}}\ \emph
  {et~al.}(2021{\natexlab{b}})\citenamefont {{Das}}, \citenamefont {{Fridman}},
  \citenamefont {{Lambiase}}, \citenamefont {{Stabile}},\ and\ \citenamefont
  {{Vagenas}}}]{2021arXiv211002340D}%
  \BibitemOpen
  \bibfield  {author} {\bibinfo {author} {\bibfnamefont {S.}~\bibnamefont
  {{Das}}}, \bibinfo {author} {\bibfnamefont {M.}~\bibnamefont {{Fridman}}},
  \bibinfo {author} {\bibfnamefont {G.}~\bibnamefont {{Lambiase}}}, \bibinfo
  {author} {\bibfnamefont {A.}~\bibnamefont {{Stabile}}}, \ and\ \bibinfo
  {author} {\bibfnamefont {E.~C.}\ \bibnamefont {{Vagenas}}},\ }\href@noop {}
  {\bibfield  {journal} {\bibinfo  {journal} {arXiv e-prints}\ ,\ \bibinfo
  {eid} {arXiv:2110.02340}} (\bibinfo {year} {2021}{\natexlab{b}})},\ \Eprint
  {http://arxiv.org/abs/2110.02340} {arXiv:2110.02340 [gr-qc]} \BibitemShut
  {NoStop}%
\bibitem [{\citenamefont {{Das}}\ \emph {et~al.}(2022)\citenamefont {{Das}},
  \citenamefont {{Fridman}}, \citenamefont {{Lambiase}},\ and\ \citenamefont
  {{Vagenas}}}]{2022PhLB..82436841D}%
  \BibitemOpen
  \bibfield  {author} {\bibinfo {author} {\bibfnamefont {S.}~\bibnamefont
  {{Das}}}, \bibinfo {author} {\bibfnamefont {M.}~\bibnamefont {{Fridman}}},
  \bibinfo {author} {\bibfnamefont {G.}~\bibnamefont {{Lambiase}}}, \ and\
  \bibinfo {author} {\bibfnamefont {E.~C.}\ \bibnamefont {{Vagenas}}},\ }\href
  {\doibase 10.1016/j.physletb.2021.136841} {\bibfield  {journal} {\bibinfo
  {journal} {Physics Letters B}\ }\textbf {\bibinfo {volume} {824}},\ \bibinfo
  {eid} {136841} (\bibinfo {year} {2022})},\ \Eprint
  {http://arxiv.org/abs/2107.02077} {arXiv:2107.02077 [gr-qc]} \BibitemShut
  {NoStop}%
\bibitem [{\citenamefont {{Das}}\ \emph
  {et~al.}(2021{\natexlab{c}})\citenamefont {{Das}}, \citenamefont {{Fridman}},
  \citenamefont {{Lambiase}},\ and\ \citenamefont
  {{Vagenas}}}]{2021arXiv211101278D}%
  \BibitemOpen
  \bibfield  {author} {\bibinfo {author} {\bibfnamefont {S.}~\bibnamefont
  {{Das}}}, \bibinfo {author} {\bibfnamefont {M.}~\bibnamefont {{Fridman}}},
  \bibinfo {author} {\bibfnamefont {G.}~\bibnamefont {{Lambiase}}}, \ and\
  \bibinfo {author} {\bibfnamefont {E.~C.}\ \bibnamefont {{Vagenas}}},\
  }\href@noop {} {\bibfield  {journal} {\bibinfo  {journal} {arXiv e-prints}\
  ,\ \bibinfo {eid} {arXiv:2111.01278}} (\bibinfo {year}
  {2021}{\natexlab{c}})},\ \Eprint {http://arxiv.org/abs/2111.01278}
  {arXiv:2111.01278 [gr-qc]} \BibitemShut {NoStop}%
\bibitem [{\citenamefont {{Chemissany}}\ \emph {et~al.}(2011)\citenamefont
  {{Chemissany}}, \citenamefont {{Das}}, \citenamefont {{Farag Ali}},\ and\
  \citenamefont {{Vagenas}}}]{2011JCAP...12..017C}%
  \BibitemOpen
  \bibfield  {author} {\bibinfo {author} {\bibfnamefont {W.}~\bibnamefont
  {{Chemissany}}}, \bibinfo {author} {\bibfnamefont {S.}~\bibnamefont {{Das}}},
  \bibinfo {author} {\bibfnamefont {A.}~\bibnamefont {{Farag Ali}}}, \ and\
  \bibinfo {author} {\bibfnamefont {E.~C.}\ \bibnamefont {{Vagenas}}},\ }\href
  {\doibase 10.1088/1475-7516/2011/12/017} {\bibfield  {journal} {\bibinfo
  {journal} {\jcap}\ }\textbf {\bibinfo {volume} {2011}},\ \bibinfo {eid} {017}
  (\bibinfo {year} {2011})},\ \Eprint {http://arxiv.org/abs/1111.7288}
  {arXiv:1111.7288 [hep-th]} \BibitemShut {NoStop}%
\end{thebibliography}
\end{document}